# Context Collapse: Barriers to Adoption for Generative AI in Workplace Settings

Emanuel Moss*
Intel, Inc.
University of Virginia

Elizabeth Watkins†
Data & Society Research Institute

Christopher Persaud
Intel, Inc.

Dawn Nafus
Oregon State University

Passant Karunaratne
Northwestern University

Mona Sloane
University of Virginia

## Abstract

As generative AI technologies are pressed into service in workplace settings, current approaches to account for the contexts in which such technologies are used fall short of users' expectations and needs. This paper empirically demonstrates, through expert interviews, both how these tools fail to account for users' context and how users deploy concrete strategies address such failures. The paper analyzes how context is variously conceptualized by tool developers, users, and social scientists to identify specific pitfalls inherent in computational approaches to context. Multiple distinct contexts tend to collapse into one another or rot, degrading over time, reducing the utility of any efforts to account for context. The paper concludes with a provocation to shift from an indiscriminate collection of context-relevant data toward a more interactional set of practices to embed GenAI systems more appropriately into users' contexts of use.

## 1 Introduction

As generative AI (GenAI) platforms expand their reach to more users, they are increasingly promoted as general-purpose tools capable of performing a wide range of tasks [32, 111]. Their ability to perform increasingly well on both general knowledge benchmarks and special-knowledge benchmarks is well-acknowledged [31] even if the evaluation ecosystem is benchmarking inapt tasks and often incorporated into training data [80]. As the capabilities claimed by GenAI platforms has expanded, there has been a corresponding interest in ensuring that these tools work well for specific tasks, producing reliable outputs germane to the individual users' contexts [89, 112]. A great deal of GenAI's capabilities come from being able to process linguistic tokens as lengthy, complex textual passages [15] and produce outputs that are relevant to the input. However, producing outputs that take into account aspects of specific users' contexts, which frame their inputs in ways that matter for producing more specific, less generic outputs remains a fundamental challenge. This challenge is beginning to materialize just as developers of GenAI are being pressed to show their tools produce favorable returns on investment. Initial studies are showing that this return is currently underwhelming, and negligible in some cases [29]. In such a setting, the ability of GenAI tools to meaningful engage with users' disparate contexts appears as an urgent concern for broader adoption of tools that meet end-users' needs. This paper characterizes the challenge of matching GenAI outputs to users' contexts by asking (A) what role users' contexts play their interactions with GenAI and (B) whether GenAI developers' attempts to utilize this context are likely to solve the challenges that face users, in ways that lead to more successful adoption of GenAI technologies?

Keystone work on GenAI has emphasized how learning statistical patterns of linguistic tokens by paying "attention" to their relative positions within the context of longer texts is "all you need" for dramatically improving model performance against natural language benchmark tasks like question-answering or summarization [115]. These so-called attention models, however, require massive amounts of text, audio, and visual training data, which developers have scraped [26], licensed, and synthesized as broadly as possible [116, 130]. Trained on massive datasets, attention models become the large language models (LLMs) and large multimodal models (LxMs) that undergird the GenAI tools and platforms that are now being hyped [108] as tools for the production of economic value through the expansion of worker productivity, and an overall tool for hyperefficiency in the workplace. GenAI tools are being pressed into service across domains as varied as medicine [48, 67, 96, 129], sales and marketing [35, 55], education [53, 84], entertainment [18], software development [99], and graphic design [49, 97], which sustains questions of how a single tool can be effective for such diverse purposes. In partial answer to this question, some researchers have even argued that "data is all you need" to improve these models' performance for domain-specific benchmarks like silicon design engineering [30].

The consequences of this are many: the risks to copyright and privacy from unscrupulous data collection [124, 126], the environmental costs of training and making inferences using ever-larger models [74], and the risk of "model collapse" from increasing reliance on synthetic outputs from foundation models for subsequent training [127]. However, as this paper suggests, some of the greatest consequences of this more-is-all-you-need approach for everyday users of foundation models are the frustrations it produces for them. User interactions with GenAI all too often produce results that are generic [112], hallucinatory [16, 113], or persistently off-topic [26, 40, 87]. Models trained on massive volumes of data can be biased toward more widely prevalent representations and steer responses away from more niche topics. Because of how models are tuned, they often--sometimes sycophantically [106]--return responses to prompts even in the absence of any grounding for their response in training data [59]. This makes it harder for users to steer GenAI systems toward what they actually need in response to their prompt within the context of their own professional needs, relationships, or clearly expressed requirements.

---

*Corresponding Author: emanuel.moss@intel.com
†Work done while at Intel, Inc.

Developers' solution to this problem has been to collect as much additional information as they can about users' interactions with GenAI tools—particularly relevant documents owned by their organization, their past prompting histories, files uploaded to chat interfaces, and metadata (time, location, browser data, device battery level, etc.), under the assumption that this additional data can serve as a proxy for users' broader context in which their need for GenAI outputs takes shape. Operating under the motto that "context is all you need" [46, 58, 128], developers have been collecting more of these types of data, not only in an attempt to deliver more useful outputs, but also to create a protective "moat" around their platforms in the hope that users would experience decreased performance upon switching to an alternate platform that did not have as deep a history of computational context data for that user [14, 19, 56].

Given the reliance placed on computational context as an "all you need" solution for model improvement, it is crucial to examine where and how such approaches fall short in meeting the needs of users within their professional contexts, and how such approaches interfere with producing the promised efficiency gains for expert practitioners. In our study, we interviewed a total of n=15 professionals using GenAI within their work contexts in which they possess significant domain expertise. Our study sampled various expert domains, including medicine, law, software design, semiconductor design, 3D design, artistic design, animation, and music production. Participants discussed how they use GenAI in their workflows, what frustrations they have experienced in ensuring outputs are both accurate and useful for their purposes, and the workflow modifications they have developed to mitigate GenAI failures. We also asked participants to engage in GenAI prompting exercises to demonstrate their prompting practices, workflow interventions, and how they made use of GenAI outputs within their broader set of work practices (i.e., what they did with outputs from GenAI systems once they produced them). Our study reveals a set of mismatches between computational context data and the social constitution of users' professional contexts. This mismatch, we argue, is at least in part produced by technical approaches deployed by GenAI developers in an attempt to account for users' contexts. The result of this mismatch is GenAI tools that only require additional work from users to reach their desired goals, potentially limiting the efficiency gains of these tools and affecting adoption of GenAI platforms.

## 2 Background

## 2.1 Overview of Context for Large Language Models

For developers of LxMs "context" pertains to multiple concepts, including the model itself, to the prompt submitted to the model to generate an output, and to the user of a model. Within a model, context refers to the region of high-dimensional vector space that represents linguistic tokens and to the text surrounding tokens in the original training data [41]. Context also refers to the data that accompanies an input to an LLM and is used in generating a relevant output [71]. Within a model, then, the context of a word like "bread" is the statistical relationship that word has with every other word used in relation to it within the training data. But in a prompt, context is also the other words in a prompt asking a language model to, for example, translate the word "bread" into German.

Developers have been working to expand the "context window" for prompts so that more, presumably relevant, information can be added to user requests [13]. This effort has led to the popularity of retrieval-augmented generation (RAG), which queries a database for relevant documents that can be appended to user inputs as additional context for an input [52, 69], allowing models to access proprietary information like company HR policies or institutional style guidelines that could not be learned from crawling the open web [11]. In addition to expanding the context window and querying documents with relevant contextual information, developers have also been attempting to collect more information about users to append to their requests, for example by collecting online behavior or the timing of online meetings and calendar events [88, 119] to append to prompts in an effort to provide outputs more useful to users. Developers have also been testing ways to equip LxMs to query users themselves to acquire needed missing information [70]. Context for developers, then, consists of data that is specific to the user's request and that can be used to refine outputs from language models.

## 2.2 Overview of Professional Context

Workplace settings require a different account of "context" than discussed above (Section 2.1). In the professions, context refers to different projects, deliverables, or clients, as work is increasingly organized in ways that require employees to make parallel contributions to multiple projects simultaneously [110]. Furthermore, organizational sociology views much of the value produced by workers as arising from a set of practices that consist of reconfiguring materials (files, documents, artifacts, etc.) and competencies (analysis, interpretation, strategy, etc.) across discrete silos in subtle ways [90]. While there have been a number of studies on how professional workflows are changing to make use of GenAI (e.g., [95], but see [6] for a thorough survey), the question of how GenAI tools adapt to dynamic work contexts or produce additional friction for workers in professional settings remains underexamined.

Computational concerns with professional context, generally, can be traced back to the work of Philip Agre, who was broadly concerned with how the use of computing shapes and reshapes professional practices as computational tools are introduced [3, 4].

Subsequently, an expanding field of study has been devoted to how workers make use of computational tools in both individual and collaborative settings, often focusing on the bridging both social and technical needs [2]. Professional context and computational practices have also been a primary concern for computational privacy [92, 93], a capacious research area that has also informed big data research ethics [131], fairness [122], and accountability [51, 107]. Beyond the frame of contextual integrity, a range of approaches to fairness [76], accountability [36], and transparency [94, 109] stress the importance of understanding how these concepts manifest within specific social contexts and how stakeholders make decisions about how to use computational tools in ways that depend on social position and relationships [28, 64].

## 2.3 Overview of GenAI Use in Professional Practices

Since the introduction of GenAI tools like ChatGPT in 2022, myriad studies have examined how they might provide value in professional settings [7, 27, 57, 68] even while they also produce significant anxieties about the future of labor [1, 18, 61, 63, 125]. At the same time, there have been significant pushes by GenAI vendors to spur adoption of these tools as part of workplace IT solutions that have translated into widespread adoption by corporate IT, even if the promised efficiency gains among workers have been halting at best. Reports about efficiency gains [21] are countered by reports of how much more difficult employees find it to work with AI-generated documents from their coworkers [91]. As GenAI has become a platform for workplace tool development [37], workers have variously been encouraged to explore and adopt widely available GenAI tools on their own, steered toward licensed GenAI platforms specifically approved by corporate IT departments, and custom GenAI tools built for internal use [123]. The specific tools professionals encounter at work and in their personal lives can therefore be idiosyncratic, shaped by the specific needs of individual professions, personal preference, and corporate vendor relations. This is evidenced, in part, by the tools participants in this study report using in their professional lives (Table 1).

## 3 Research Questions

As GenAI tools proliferate in the workplace, developers have emphasized the importance and benefits of accumulating more context data from which to generate better outputs for users, thereby increasingly the computational efficiency GenAI tools and, by they hope, the efficiency of users [46, 58, 128], engaging in a set of emerging technical practices that coalescing under the label of *context engineering*. However, any efficiencies gained by the accumulation and use of context data depend on existing workplace practices and how well new tools are incorporated into existing workflows. To investigate the potential of accumulation of context for developing GenAI tools that can increase workplace efficiencies, our research questions are:

- What prompting practices do professionals use when interacting with GenAI tools to produce outputs from these tools that are relevant for their workplace context(s)?

| GenAI Tools | Non-GenAI Tools |
|---|---|
| ChatGPT | Clipboard History Pro |
| Claude | Overton Policy Commons |
| DoximityGPT (OpenAI) | Semantics Scholar |
| CoPilot | OneNote |
| Grok | Zotero |
| Gemini | Microsoft Office Suite |
| Open Evidence | Photoshop |
| UIPath | Procreate |
| NotebookLM (Google) | Fresco |
| Uma (Upwork) | Google Suite |
| Stable Diffusion | Nuance Dragon |
| Midjouney | Maya |
| Ideogram | Blender |
| 11 Labs | Canva |
| Runway | Figma |
| Quso | Unity |
| Firefly | Adobe Suite |
| Luma | GraphPad Prism |
| Tripo | JMP |
| MetaHumans | Lab Collector |
| Convai | Strand NGS |
| Shotgrid | Osirix |
| Bonzai | OnePacks |
| AppWiz | |

Table 1: GenAI-enabled and non-GenAI-enabled tools used by participants (n.b., GenAI is increasingly integrated into productivity software, e.g., Google Suite. Also, non-GenAI tools may include more classic ML techniques like NLP or ASR).

- What lessons do the prompting practices and workplace contexts of professionals using GenAI hold for context engineering?

## 4 Methods

The research questions listed in Section 3 arise from an exhaustive period of desk research on the divergent understandings of context in social sciences and engineering, some of which are discussed as findings in Section 5.2.1. To address how *users* address context with respect to GenAI tools, the research team engaged in a series of semistructured interviews with individuals who self-identified as professionals who used GenAI to complete their work on a regular basis. The research study was reviewed and approved by an internal process to ensure the privacy and safety of research participants, as well as to identify and mitigate any research ethics concerns. All interviews were conducted on a voluntary basis, were centered on the workplace activities of research participants, and were incentivized by a nominal monetary award.

## 4.1 Recruitment Process

Participants in this study were recruited through a user-studies platform[1] that sources voluntary participants who self-identify as meeting researcher-defined demographic criteria. These criteria recruited for adults of any gneder over 18 years of age or older from the United States or Canada who had completed an undergraduate degree, a postgraduate degree, or a vocational or trade school, and who had either fluent or advanced English language proficiency. Additionally, potential participants were asked:

(1) Do you use AI tools? (Options: Yes, No)
(2) What AI tool(s) do you use? (Options: ChatGPT, Gemini, CoPilot, Claude, Grok, Stable Diffusion, Other)
(3) How often do you use AI tools for important tasks, i.e., work tasks, serious hobbies, or to help make important decisions? (Options: Often (Every day), Sometimes (Once a week or more), Rarely (Once a month or less), Never (Only use these tools for fun))
(4) What is your professional are of expertise or training? (Short response requested)
(5) What do you do for work? (Short response requested)
(6) How would you describe your attitude towards AI? (Options: Enthusiastic, Cautiously Optimisitic, Skeptical, Undecided, Neutral, Other)
(7) What do you generally use AI tools for? (Short response requested)

A total of n=235 individuals who met all demographic requirements (age, location, education, language proficiency) responded to the request for participants. Of these **n=15** participants were selected based on their professional role and frequency of AI use, optimizing for a range of roles and a high frequency of AI use.

## 4.2 Participants

Participants were chosen to represent a range of professional roles who use AI often. They are described in Table 2. Of the **n=15** participants, 12 identified as men, 3 identified as women, 11 identified as white, 2 identified as Black, 2 identified as Asian, and all had either undergraduate or postgraduate degrees.

## 4.3 Interviews and Demonstrations

For data collection, the research team employed a modified semi-structured interview research protocol [23, 100] comprised of three sections. The first section of the interview protocol consisted of standard questions that were posed to each research participant. These questions were:

- How old are you?
- Could you tell us a bit about your general background, including your education?
- What AI products or services do you use?
  - If you use more than one, what do you use each one for?
  - Do you use one more than another?
  - Do you have a sense for waht one tool is better or worse for than another?

| ID | Professional Role | Age | Income Range |
|----|-------------------|-----|--------------|
| P01 | Graphic Artist | 46 | $30-40k |
| P02 | Corporate Research Analyst | 43 | $150-200k |
| P03 | Web Content Writer | 63 | $30-40k |
| P04 | Certified Public Accountant | 42 | $150-200k |
| P05 | Physician | 31 | $100-150k |
| P06 | Legal Manager | 52 | $100-150k |
| P07 | Rsearch Project Manager | 37 | $60-70k |
| P08 | Dog Trainer/Musician | 32 | $80-100k |
| P09 | Pharmacist | 38 | $100-150k |
| P10 | Event Organizer | 56 | $60-70k |
| P11 | Biological Scientist | 33 | $200k+ |
| P12 | Microbiology Technician | 31 | $30k or less |
| P13 | Radiologist | 57 | $200k+ |
| P14 | Graphic Designer | 57 | $80-100k |
| P15 | Animator | 32 | $60-80k |

**Table 2: Research Participants**

The second section of the interview protocol included more open-ended questions intended to elicit descriptions of actual interactions between participants and GenAI tools. It should be noted that since the introduction of ChatGPT in late 2022, "AI" has come to mean "GenAI" within popular discourse, and therefore no such distinction is made within the interview protocol, however any distinctions drawn by participants was explored through follow-up questions. Each of the questions listed below would be followed up with requests for additional details, screensharing of results (where appropriate and permissible), or demonstrations. Not all questions were asked to all participants, but responses included rich descriptions of prompting practices, experiences, and value added (or not) to professional tasks.

- In general, what do you use AI for? (continue asking until all uses have been described)
- Is there anything you would never use AI for?
- Can you tell me about the first time you recall doing something with AI that made you think it could be useful professionally?
- What is the most frustrating interaction you recall having with AI?
  - What was frustrating about it?
  - What do you think that tells you about how these tools work?
- What is the most delightful or surprising interaction you recall having with AI?
- What is the longest sustained interaction, conversation, or thread you recall having with AI?
- For the various tasks you use AI for, how does your time divide across those tasks?
- How would you rate your skills at using these tools?
- What have you learned, in terms of how to use these tools, that has made them more useful?
- What are your sources of information for how to use these tools?

- Is there anything you thought you would be able to use these tools for that has never quite worked out the way you expected it would?

The third section of the interview protocol consisted of an exploratory exercise in which participants were asked to demonstrate how they would go about accomplishing a professional task. A total of 11 out of 15 participants volunteered demonstrations or screen-sharing of past tasks, while four engaged in prompting demonstrations upon request.

### 4.4 Analysis

Interviews were recorded and transcribed using a qualitative analysis platform[2] that employed automated speech recognition (ASR). Transcripts were reviewed alongside audio-video recordings, and any errors in ASR were corrected by the research team. The transcripts were analyzed using a grounded theory approach [54] that produces knowledge through inductive reasoning [22] to generalize higher-order concepts from disparate evidence that are likely connected through those higher order concepts. The grounded theory approach undertaken in this study consisted of three phases of iterative, inductive reasoning.

The first phase consisted of labeling sections of interview transcripts using an open coding technique. Participant quotes that were likely relevant to the research questions were identified and labeled with simple descriptors. A total of n=59 "open" codes were assigned in this phase, and included labels like "professionalism", "context switching", and "frustration". By the open nature of coding in this phase, these codes were idiosyncratic and occasionally duplicative. This shortcoming was addressed in the second phase of analysis: axial coding [17]. In this phase, open codes were grouped into related categories, which were then given labels and provisional definitions. The axial codes and their definitions served as the codebook for the third phase: selective coding. During selective coding, interviews were re-analyzed and coded according to the higher-order labels to extract relevant quotes and refine the concepts found relevant to answering the research questions given above. The codebook used for selective coding can be found in Table 3.

### 5 Findings

#### 5.1 Using GenAI in Workplace Contexts

The semi-structured interview format, with opportunities for demonstrations, elicited numerous instances in which participants described how they use GenAI in workplace contexts. The interviews elicited n=29 descriptions of actual prompts used by participants in their work (see Appendix 10), as well as descriptions of their general approach to prompting (n=22). The interviews also elicited insights into what aspects of their workplace context are relevant to their interactions with GenAI (n=22) and a limited number of explicit descriptions of how they tactically set the context of their interactions with GenAI (n=8). Additional relevant qualitative labels applied to the interview transcripts are contained in Table 3, with short descriptions from the code book employed by researchers, as described in Section 4. These codes were used to label quotes from participants that were key to the findings presented below. Exemplary quotes that illustrate the claims made in each section of Section 5 are indicated in-text and can be found in referenced tables. All quotes referenced are included in Appendix B.2.

| Label | Definition | Count |
|---|---|---|
| AI Tasks | A mention of a task accomplished with GenAI | 29 |
| Aligned to Context | A mention of a result that was appropriate for the context, as it was understood by the participant | 28 |
| Prompt | An explicit description or example of a prompt | 24 |
| Aspect of Context | An aspect of participant's work that was relevant to how they understood the context of their requests to a GenAI system | 22 |
| Prompting Practice | A description of how the participant used GenAI tools. | 19 |
| Multiple Correct Answers | A description of an interaction with a GenAI tool that did not have a single 'correct' answer, and for which 'accuracy' was not the most relevant measure of success. | 17 |
| Egocentric | An evaluation of a GenAI output, the reference point of which was the participants' own subjective judgements or feelings | 17 |
| Adjectival Play | The use of adjectives in prompts to steer GenAI outputs | 14 |
| Trouble | An instance in which the interaction with GenAI caused trouble for the participant. | 10 |
| Tone | An instance in which the participant was concerned with the style or tone of an output from a GenAI system | 9 |
| Good Enough | A description of an interaction with a GenAI tool that resulted in an ouput that served a purpose, either because the stakes were low or the output could be further revised outside the GenAI tool. | 8 |
| Context Setting | An explicit description of how a participant would attempt to control the context of their interactions with a GenAI tool | 8 |
| Privacy | A mention of privacy concerns, in which privacy limited the information the participant was willing to enter into a prompt | 6 |

**Table 3: Codes and definitions used in codebook for selective coding.**

##### 5.1.1 What counts as context for GenAI users. Participants in this study had various understandings of what constitutes the context of

their needs, and how those needs should be communicated as inputs to GenAI (see Table 4). Documents or files comprise a significant dimension of context inputs. For some [P02, P05, P07], context is at least in part set by a collection of documents they intentionally provide to the GenAI platform. These participants provide patient notes [P05:Q03] to generate personalized release notes, meeting notes [P07:Q04] to generate meeting summaries for distribution to attendees, and personal writing [P05:Q02] or favorite novels and essays [P02:Q01] to generate outputs that are unified in style and tone. Other participants [P09, P11] using proprietary tools operate under the assumption that outputs rely on internal documents to produce outputs [P09:Q05] but were not necessarily aware of which documents the tools they used had access to [P11:Q06].

Context, for participants, was not entirely document-based (see Table 5). At least one participant sees individual threads within a GenAI platform as constituting separate contexts [P07:Q07]. Others assign specific roles, either to the GenAI tool or to the intended audience, to situate an interaction within a general social context [P05:Q08, P05:Q15, P09:Q09, P10:Q23] or a specific one [P07:Q10, P05:Q14]. Commonly, partipants use themselves as the contextual frame for an interaction, seeking outputs that are useful or make sense "to me", saying explicitly "I am looking for tools that can act in the same way that I act" [P02:Q11]. This sentiment was also frequently expressed less explicitly, when participants would use descriptors ("tourist trap"[P06:Q12] or "walking distance"[P10:Q13]) that have idiosyncratic but deeply personal meanings to each participant. At times, participants understandings of context map onto computational approaches to context: stores of documents, previous interactions within a thread, instructions explicitly delivered to a GenAI platform. At other times, it varies from what constitutes context from a computational perspective: managing conversations between multiple other interlocutors, signaling distance from a desired outcome using paralinguistic cues (e.g., using filler words like "um" or "er") [81], or using the self as a reference point for taste-based and spatial considerations [38].

### 5.1.2 User strategies for managing context.
Participants in this study also revealed strategies they employ for managing the context of their interaction with GenAI in ways that steer outputs toward what they need in their personal and professional lives (see Table 6). These strategies consist of:

- Managing threads
- Aiming for the "good enough"
- Getting to a starting point

Actively managing conversational threads within a GenAI platform was a key strategy. Managed threads could correspond with discrete topics, projects, or clients [P07:Q07], as discussed above (Section 5.1.1). But the strategy of managing threads also includes constraining what users' ask of a GenAI platform to stay within the bounds they trust are relevant to their goals. In hardware design and software engineering, studies have shown this looks like breaking down projects to atomized components that can be hand-inspected for errors [87]. In other fields like legal research or accounting, this looks like requests for outputs that are "good enough" to serve as a starting point for additional work (see also [20, 86]). One participant describes their technique of asking for an 8 sentence summary, from which they craft the 2-3 sentence summary they actually need

| **Q01:** "For GenAI, I have NotebookLM. I have 40 books that are really influential to me that I have in NotebookLM and then I'll query against that. I also have my conversation history on a question and answer forum so that I can go back and I can like have chats with myself and, and look up things that I've written in the past in terms of the stuff that I have that I use on a regular basis." [P02] |
|---|
| **Q02:** "... if you say to ChatGPT store in your long term memory, whatever, it will lock that into its long term memory. And so I have pasted a ton of stuff that I've written into it. And so I had to respond to this, you know, this like silly interview the other day for the psych department about like, oh, we're going to feature you on Instagram. Answer these questions. I did not have time to do this. And so it has read hundreds of pages of my writing and it stored it in long term memory. So I said just respond to these questions, knowing what you know about me. And it responded two sentences to each of them. It sounded exactly like me, you know, and it was an appropriate answer to all those questions. I made minimal changes." [P05] |
| **Q03:** "So I use that tool [Doximity] all the time at work to generate documents. For instance, if I need to generate a hospital course summary, I can paste in maybe 20, 30 pages of information from someone's hospital course, their admission, HMP history and physical, their most recent progress note, most recent consultant notes, and then it will generate a timeline of exactly what happened while they were in the hospital, when this MRI came back, that kind of thing." [P05] |
| **Q04:** "I use it for taking like a lot of my thoughts or like my random like notes and just like typing it all to chatgpt. And I say like make sense of this and put it into a document and it kind of gives me a baseline for a narrative that I typically usually have to kind of flesh out a little bit more, make it a little bit more specific or more detailed to the actual event or whatever I'm reporting on. But it's a really good way to just get... like I have all of these flip charts from this event, I have all these sticky notes from the event. They're all typed up. So like let's get it into like a readable format, like four pages maybe." [P07] |
| **Q05:** "The system like the CCE tool can pull information, all these files for me that I need, save me time from having to look through all those documents. So it highlights information I need to be able to pull and complete the case. And we have a system that automatically generates what we call criteria. The criteria is the form that's used to get whatever drug is being requested, approved or denied. So there's a system that populates the questions based off the answers that are provided. And that's a system that's, you know, internally built by the company. I don't know what is called cast, but each specific insurance plans guidelines are in there, we have to follow that." [P09] |
| **Q06:** "**Q:** Okay, so the company specific ChatGPT that you mentioned that [vendor] built, does it have access to proprietary information? Can you ask questions about your own products and get some answers? **A:** Yeah, I think so, yes. So the because it's company's products so we are free to type in whatever we want." [P11] |

**Table 4: Quotes from participants expressing context as document-based.**

**Q07:** "I love it for briefing documents. Master's programs are all reading and I don't have time to read all of it, so it'll give me really good overview for the most part, at least enough to get through class, which is exactly what I need. I've had it, like, put things in tables for me so I can see it in a different way." [P07]

**Q08:** "If I'm writing patient instructions, I want it usually at the fifth or eighth grade level. But if I'm writing a discharge summary hospital course that is not for patients... it should be written in formal technical language that is accessible to those other physicians that are reading this. So that's why I put that information in there." [P05]

**Q09:** " I'm meaning to include this specific rationale for denial in this but ... make it to where even like a fifth grader could read it. Like ... be specific with the prompt because you have to be specific to get like a specific answer. But put in stuff like that that will be able to make it understandable for the patient." [P09]

**Q10:** "It's like a family group chat. My mom started it, and he [my nephew] was really going against her. And... this was the one time I could actually ... defend my mom, which is ... rare. So I would like, copy, what he said on my phone and then put that into ChatGPT and say, how should I respond to this? And then I would copy it and put it back. And then, yeah, just kind of went back and forth." [P07]

**Q11:** "I am looking for tools that can act in the same way that I act." [P02]

**Q12:** "Get me out of a tourist trap, find me a local experience, please. I'm going to be very plain language, not tourist trap estaurants. Include unique, hole in the wall, places that exceed expectations." [P06]

**Q13:** "Mostly I just wanted to see what was in walking distance from there. So that would kind of narrow my plans down of what we're going to do is something that would, you know, be close enough to walk." [P10]

**Q14:** "But with Open Evidence it's just simply a search engine that gives you specific tailored evidence based results that pull from papers. And so it's basically just a faster way of searching PubMed using advanced search... The attending would ask ... my intern: With their CKD [chronic kidney disease], can we start Entresto? And and they would go, I don't know. And I'll go, listen you guys, move on to the next system, I'll look this up on Open Evidence. And then they finish talking about, you know, what we're going to do from a protein calorie malnutrition standpoint, which, say, another problem the patient has. And by the time that's done, it's been 30 seconds and I've pulled this up. Okay, you know what? I think it is fine to start that. Let's wait for today's echo to come back, and then we'll go from there. So that's how it would work in that flow." [P05]

**Q15:** "I've trained my private one to be extremely, like, you know, like, loose with me and very blunt and honest and funny. And it will be a bit irreverent. And so it will be like, you know, 'Yo, Captain [Lastname]!'." [P05]

**Q23:** "So I would input, I would just say I need an email written to the executives to. A professional email to executives inviting them to this event and yada yada. And so that's how I used it at work most of the time, was to make myself sound more professional." [P10]

**Table 5: Quotes from participants expressing themselves or an imagined reader as a contextual frame.**

**Q07:** "I love it for briefing documents. Master's programs are all reading and I don't have time to read all of it, so it'll give me really good overview for the most part, at least enough to get through class, which is exactly what I need. I've had it, like, put things in tables for me so I can see it in a different way." [P07]

**Q16:** " I find that eight sentences gives me enough information that I can edit it down to the two to three sentences that go in the newsletter for each of these individual stories." [P02]

**Q17:** "It needs to be helpful, useful, informative. I'm sure the customer reads it when he gets it and looks it over to see if there's any serious problems. They can ask for a revision, although nobody ever does rarely anymore. And they can go in, maybe change some things too if they want to." [P03]

**Q18:** "I adjust it to my style. I always still tweak the final response. I take word processor or something. Yes, I will take it over to word, you know, because I always [personalize it]." [P06]

**Table 6: Quotes from participants expressing themselves as a contextual frame.**

within the context of a newsletter [P02:Q16]. Another described using GenAI for clients who were less likely to closely scrutinize work submitted to them or be bothered by any necessary corrections [P03:Q17] while another always edits every output anyway, to preserve his own voice [P06:Q18].

Participants also manage how well GenAI outputs interact with their context by scoping requests to topics that do not have a single correct answer (see Table 7). Similar to requesting "good enough" outputs, such interactions move work forward without requiring additional validation or fact checking. Summarizing and reformatting information in different, arbitrary ways to aid in learning a topic, is one example mentioned by participants [P07:Q19]. Other areas where there are multiple correct outputs include requests for lesson plans, where any variation on a theme will suffice [P04:Q20], or in brainstorming where a starting point and not the finished product is the goal of the GenAI interaction [P06:Q21]. Participants also used GenAI tools in ways similar to brainstorming when they did not know exactly where to start, either in something as challenging as making a diagnosis from an MRI [P13:Q22] or as simple as writing a Valentine's Day card [P07:Q24].

## 5.2 Lessons for Context Engineering

Having established user strategies for bounding the context of their interactions with GenAI, we can compare their strategies with developers' engineering approaches for incorporating context on behalf of users. We do so to answer our second research question about how computational approaches to context affect professional work practices with respect to GenAI by pointing to a mismatch that appears between what engineering approaches conceptualize as context and what users need from context. Because it depends on our findings presented in Section 5.1, this mismatch is introduced conceptually below in Sections 5.2.1 and 5.2.2, and then our findings relevant to this mismatch are presented in Section 5.3.

*5.2.1 Developer strategies for context engineering.* Only partially overlapping with how GenAI users understand and manage the contexts in which they use GenAI tools, engineers have focused their attention on advancing the means through which data can be added to users' prompts through larger context windows, through RAG databases, through lengthy memory, and through online activity logs. It is no surprise that there is an increasing effort to expand the data that can be served as context for users' prompts, and to serve it more effectively. This effort has become known as *context engineering* [82]. The promise of context engineering is that those who operate models can accumulate data extensively as a way to differentiate their products from competitors, and use context engineering to provide a higher quality of outputs to their users. For major GenAI vendors, the data they call "context" can serve as a moat around their products, ensuring that anyone contemplating shifting to another product will see lower performance elsewhere [14]. This thinking transforms "context" from additional data used to improve an output to something that can be hoarded, and from which value can be extracted for the platform and its users.

| |
|---|
| **Q19:** "I love it for briefing documents. Master's programs are all reading and I don't have time to read all of it, so it'll give me really good overview for the most part, at least enough to get through class, which is exactly what I need. I've had it, like, put things in tables for me so I can see it in a different way." [P07] |
| **Q20** "Another one that doesn't have much back and forth is I've used it sometimes just to help teach me a topic. Sometimes it's teaching me a topic that I know, but I want to see how it tells me so that then maybe I can try teaching someone else in a different way. Like, like I understand this, but I need some better ways to include it in a presentation. So I've done that." [P04] |
| **Q21:** "A lot of times I'll use it as a brainstorming tool. And I know that when I brainstorm, I'll say, rewrite the following paragraph in first person in the same tone in paragraph form and any other things." [P06] |
| **Q22:** "But sometimes you just want ChatGPT or you can put things in. I'm always hesitant 'cause. I'm like, should I trust it? But at least it's a starting point. Then I can say OK. This sounds reasonable. Let me go look that up. Look up some of the suggestions now in STATdx or something and see if that makes sense. And often that will have what's called a differential diagnosis, where, well, the adrenal adenoma.... here's other things it could be. Look for these findings and things, and so it gives you somewhere to start. If you're having trouble, like getting your feet on the ground to know like. OK, what is this? I've never seen this, never saw this in residency. Never seen a case like this. There's no history to tell me what. Strange sarcoma. Whatever tumor this guy has like. So sometimes it can help you like narrow things down, or at least get started on a search. And that's probably where I use that kind of stuff the most." [P13] |
| **Q24:** "For Valentine's Day, I was having [my partner's] car detailed inside and out, and I wanted to make him a card with, like, a riddle. And so it kind of like, gave me a riddle for him to start to guess. And I said, can you make it a little bit longer? And then it made it a little bit longer. And then I said, do you have an alternate version? I was just curious. And then they gave me an alternate version. And I must have went with that or the one before it. And it was just like, you know, I just wanted something silly because, like I wasn't going to be able to give him the car detail that day." [P07] |

**Table 7: Quotes from participants discussing tasks where an ouput can be 'good enough' without being steered precisely to a relevant context.**

The problem with each of these approaches is both technical and social. Technically, as measured by widely-accepted benchmarks, adding more context does not necessarily improve performance. In fact, as context windows grow, LLM-based approaches are less likely to retrieve useful information from the middle of the added data within that window [44, 72]. Long context windows can ironically degrade performance, as the larger data quantity becomes increasingly like an expanded haystack in which to find a needle [77, 85]. As we have seen above, especially in Section 5.1.1, what counts as context for users does not map cleanly onto what AI developers see as context. There are recent developments that depart from this general rule, including Anthropic's "memory for teams" feature that allows project- and team-based contexts to be explicitly declared [9]. In general, however, for users context is not an additional set of documents for grounding outputs, it is something else entirely.

### 5.2.2 Social understandings of context.

To understand what context is for users, as revealed within this study, and the tension with how context is operationalized within context engineering, requires a brief detour through anthropologist Nick Seaver's observation that "the nice thing about context is that everybody has it" [105]. By this he means that context should not be seen as something present or lacking, where you can add more or less of it, but instead a decision about what counts as context. He shows that "context cultures" shape how people arrive at determinations of what counts as context. In some ways, today's context engineering follows a very old "context culture" of context-aware computing, a staple of mid-2000s ubiquitous computing. That paradigm took context to be a set of stable inputs like that can be easily delineated from other data, like location or user identities, and left on the table the possibility of defining context as an outcome of interactions, whether between people or people and machines, that comes into being when the protagonists involved mutually recognize it as a context [43].

This interactional definition of context is one that could be considered a staple across many social scientific subfields, but is comparatively underdeveloped in computational approaches, particularly those described in Section 5.2.1. In Sections 5.1.1 and 5.1.2, however, we saw demonstrations of users' interactional understandings of context, wherein shared framings are formed as the result of a person-to-person or person-to-machine interaction. This creates a contrast that helps reveal the limitations of currently emerging approaches to context engineering.

## 5.3 Limitations of context engineering.

The focus of context engineering (Section 5.2.1) results in hoarding a special category of data that could be called *context data*, rather than building shared frames across person and machine that bestow meaning on data. This produces two key pitfalls for approaches to context in GenAI development: *context rot* and *context collapse*.

Below are findings drawn from contrasting social scientific and computational approaches to context, validated by reports of how such contrasts produce trouble for users of GenAI.

*5.3.1 Context Rot.* The concept of hoarding data as context, collecting it and drawing a moat around it, assumes that context data is a stable store of value, but it actually invites context rot. As anyone who works with data or data governance knows, data can become less reliable over time, either because the infrastructure that makes it useful changes [25], because the phenomena in the world that data represents drifts away from the status quo [121], or because data collection and use changes that which it measures [47, 75, 83]. This poses an even greater challenge for context data. It is far more perishable because transforming context into context data, attempting to preserve it in databases and chatlogs, tears it from the social packaging that gives it meaning. Documents added to a RAG database can be deprecated and replaced by subsequent versions, just as the motivation for a query about food allergens can be made in the context of a guest coming to dinner rather than a permanent change in diet.

One participant recalled a chatbot unexpectedly referencing an obscure, advertisement for an event from the distant past that had passed across his social media feed, as if it was directly relevant to a separate conversation [P08:Q25]. Another participant observed the tool he was using fixate on content provided in a prior conversation, regurgitating it at a later time without recognizing the older information was no longer useful in the new context [P11:Q26]. Without the additional social context of organizational change and personal relationships, context data becomes far less valuable over time, and presents its own computational challenges (see Table 8). If there are two versions of a document in a RAG database, it can be difficult to query that database in a way that returns the most relevant information, from the most relevant document. While a more recent version might be the most relevant for a query, that is not always the case, for example if a user is working on a legacy code base written in a deprecated version of a programming language. An expanding horde of context data makes the haystack bigger, without making the needle in it easier to find [66].

*5.3.2 Context Collapse.* There are currently very few ways to distinguish between context data points that are relevant to different social positions users occupy across their interactions with GenAI systems. These positions collapse into a diffuse, continuous space even when they are quite distinct. In this way, GenAI systems replicate a central context problem of social media platforms at scale-—users' self-presentations and inputs are refracted through the synthetic media outputs in ways that foreclose the possibility of different self-presentations for different audiences [24, 39, 79, 117]. Multiple social contexts can be relevant concurrently while remaining well-bounded, and they can succeed one another serially. A lawyer working for two different clients needs to maintain that boundary, even while working with documents that greatly resemble one another. A consultant who moves from one job to the next similarly might deploy identical strategies, but needs to be clear about what is particular to the previous client and what parts of a workflow can be replicated for the new client. Similar boundaries exist between our professional and personal lives, and there can be consequences when those social contexts collapse. However, it is not obvious how contexts can be kept from collapsing into one another once placed in a horde of context data. Even in workplace settings where there is a formal separation between personal and professional tasks, project contexts can collapse into one another. Relatedly, the onus falls on workers themselves to keep personal life completely separate form workplace tools, or risk a collapse of contexts.

| Q25: "I was talking to [a female-coded chatbot] about something, and then she referenced some sort of event that I had taken part in. And she asking me about it, but the only reference to it was some random Instagram post or flyer that came from the event. And so it was clear that she was just drawing from information that Instagram knew about me. But it didn't feel like it was a normal interaction. It wasn't like somebody that would just walk up to you randomly, not knowing you and be like, 'So how was it? Did you like this?' It was just weird." [P08] |
| --- |
| Q26: "It could be very stupid, it could like just repeat what you told it. Like you said, 'Oh, for this specific type of experiment LSA-XT can be the optimal choice'. Then the next time you asked 'Hey, do you have like new idea about this assay? Do you think I should use a different type of technology for this assay?' but it will keep mentioning LSA-XT. Because it appears you were the only one who told ChatGPT that LSA-XT is a good technique for this experiment. So it just keeps telling you this is the best." [P11] |

**Table 8: Quotes from participants discussing context rot.**

Context collapse is a problem for trained language models, apart from efforts at context engineering (see Table 9). A participant working as a radiologist reported difficulties retrieving relevant information from a GenAI system within the context of their specific patient, as the model had collapsed data relevant to a range of patients into a single context [P13:Q29]. For this radiologist, he had to wade through outputs suggesting information about pediatric and uterine diseases that were irrelevant to his adult, male patient. Approaches to context engineering require ongoing curation of context data through the collection of additional metadata, and if the categories of contexts are not made explicit prior to the collection of context data, this effort can fail. Recent approaches such as contextual embeddings, context enrichment, or context-aware querying reduce, but do not eliminate the likelihood of failure by extracting contextual clues from documents [45, 118], linking segments of files to each other to expand their contextual relevance [34], or simply adding additional data (context data inferred as relevant) to users' prompts [8]. These approaches still require a nexus to a user's context to be affective, as they do not have access to aspects of context not already represented in datasets.

Additionally, contexts that seem superficially related, at least at the level of individual tokens may be quite distinct, presenting problems for someone researching 'nuclear medicine' when a GenAI system returns data about physicists who deal with nuclear topics, but are also represented in the literature as 'doctors' [P02:Q27]. Crucially, separate social contexts are not always able to be explicitly declared. In sociolinguistics, contexts can shift [50, 120] through subtle linguistic cues or simply if another person enters the conversation. Similarly, contexts in which interactions with GenAI

**Q25:** "But with Open Evidence it's just simply a search engine that gives you specific tailored evidence based results that pull from papers. And so it's basically just a faster way of searching PubMed using advanced search... The attending would ask ... my intern: With their CKD [chronic kidney disease], can we start Entresto? And they would go, I don't know. And I'll go, listen you guys, move on to the next system, I'll look this up on Open Evidence. And then they finish talking about, you know, what we're going to do from a protein calorie malnutrition standpoint, which, say, another problem the patient has. And by the time that's done, it's been 30 seconds and I've pulled this up. Okay, you know what? I think it is fine to start that. Let's wait for today's echo to come back, and then we'll go from there. So that's how it would work in that flow." [P05]

**Q27:** "We were looking for information on doctors who had been involved in nuclear medicine in Canada by decade. The results that kept coming up when we were using generative AI included a lot of people who were physicists, not [medical] doctors, but because doctors ahead of their name. The generative AI just doesn't do that. I am always looking for something that's going to speed up my work or enable me to do things faster, but it has to be accurate. And a lot of this comes in to thinking through what generative, like what the generative AI tools are particularly good at." [P02]

**Q29:** "Sometimes [Gemini is] way off and I ... try redirecting it. Then I'll have to make a decision. Do I want to redirect it from what it gave me? OK. It's a middle-aged male. You're giving me pediatric answers and then it will reassess and I'll kind of give it a chance that way. I'll try to just give it more whatever information I have to try to guide into something else, and often it will like click into a more reasonable differential diagnosis that I can go research on real tools. That doesn't happen too often." [P13]

**Table 9: Quotes from participants discussing context collapse.**

occur can change in ways that are impossible or difficult to store as context data. If a user interacting with a system abruptly asks someone else to look over their shoulder at their interactions with a chatbot, the context has changed in ways that may or may not be relevant to the usefulness of context data. For example, we found that in hospital settings, doctors will sometimes ask a specialized chatbot for information they are unsure about when on their own, but will then use that information as part of a separate or ongoing conversation [P05:Q14]. Careful curation of metadata to prevent context data collapsing into one another is a partial, computational, solution, but in practice our research found that most of the work of context curation fell on users who must manage their threaded conversations carefully, use different tools for different contexts, or otherwise balkanize their work in ways that make their workflows unwieldy but keep their contexts intact.

## 6 Discussion

The idea that context data can be stored indefinitely and filtered to provide relevant additional details to users' subsequent prompts mistakes users' interactions with GenAI systems as taking place solely within the context of the relationship between user and system. However, users occupy multiple contexts simultaneously and users' contexts change continually. From the point of view of linguists, social scientists, and others who study the production of meaning, the contextual elements that give symbols (i.e., data) meaning are continually constructed by those who participate in a context [5, 62, 102]. Participants within a conversation, or anyone sharing a context with another, constantly negotiate the roles they play and the norms they follow, with roles and rules changing from one moment to the next. The value of context data, therefore, is lessened when it is isolated in data stores, away from the relationships in which it gave that data meaning. For example, we observed that artists working on ideation for a game universe generate numerous pieces of concept art based on reference material that includes intellectual property, however they are not permitted to carry these proprietary materials further in the production pipeline. What might be acceptable for a design phase in which colleagues are throwing ideas at each other in a private setting becomes completely forbidden when their ideas need to be shared to broader public audiences within and beyond their workplaces.

Having argued that context is what gives data meaning and after enumerating the pitfalls of hoarding context data, a significant question remains: If context is what gives data meaning, and if it is always somehow superfluous to data, how can data-driven machine learning systems ever access context? The short answer is that all manner of such systems have access to context. Systems that have been evaluated in accordance with the IEEE 7010 standard for assessing the well-being implications of artificial intelligence [101], for example, are accompanied by documentation that explicitly state their intended and unintended uses, which bounds the context in which the system is being used, giving the inputs and outputs contextual meaning. GenAI systems, too, contain exogenous sources of context. Safety guardrails [10, 42] keep model outputs within the bounds of legality and propriety, while metaprompts and reinforcement-learning based fine tuning approaches place the system in specific roles vis-à-vis the user, the relationship between which constitutes an important dimension of social context [12, 33, 114]. Browser integration for agentic AI involves multiple forms of context engagement, including providing LLM-based "agents" with browser and websearch access (such as perplexity.ai), with the presumption that reasoning abilities should adequately enable such systems to sufficiently learn from and navigate through text and multimodal web content. These processes give some level of additional meaning to data that would otherwise be more polysemous or uncertain than without such contextual framings.

However more techniques are needed to give context data the meaning it is intended to hold. Bounding context data with time frames within which it is relevant is one necessary step. context data that is intended to persist should be clearly delineated from context data that is intended to expire, and developers should pursue ways of distinguishing between these categories of context data, as current approaches rely on already-existing metadata to enact processes such as data expiry [65]. One way to do this is through the measurement of uncertainty in model outputs. Evaluative metrics for model uncertainty [60, 73, 98] can provide clues as to whether a piece of context data increases or decreases model uncertainty, and agentic approaches can leverage high uncertainty to return requests for clarification of context to users. For example, a user input that might invoke a deprecated piece of context data ("Can you recommend an investment strategy to me?") might return questions about the context data itself ("You said you were saving for a downpayment six months ago, is that still a goal I should consider as part of my recommended strategy?"). Dialogues like this map onto how context is established socially, through conversation and negotiation about matters of fact, roles, and norms. While such interactions can, in fact, be generated at the time of query, they do not seem to be implemented in any publicly-available GenAI system.

Another way of incorporating context into such systems is in evaluation. Current evaluation practices are based on static assessments of a model performance on a task, typically assessed in a vacuum. A more comprehensive approach to evaluation would integrate tools for analyzing context, not just immediate human-AI interactions but potentially even broader effects such as societal impact. Field testing, randomized controlled trials [78, 104], and qualitative social-science approaches like user studies, interviews, and real-world observations can all elicit greater insights into how systems interact with, adapt to, and utilize contextual data.

## 7 Postscript: All You Need Is not Enough

A recent popular explainer video states that "context is the complete information environment that surrounds an AI model's decision making process–everything it can access and consider when making a response". The video then calls for context engineering as an approach that explicitly designs that information environment. The emphasis here is on dynamic systems that are not "just a static prompt template [but] the output of a system that runs before the main LLM call [which generates] the right information at the right time" [103].

From a social scientific perspective, there is much to recommend such a call. The emphasis on the dynamic nature of AI interaction echoes longstanding approaches to social interaction in many subfields, whether human-to-human, human and machine, or humans and non-humans more broadly. The emphasis on selecting relevant information for a task, as opposed to determining the factuality of information whether relevant or not, also echoes recent calls to rethink model testing towards live, in-situ use [104].

However, as we have discussed, context is not a static material but an emergent outcome of interactions. Context is an emergent property of the human-machine interaction, where what constitutes the "relevant" frame of reference is negotiated, maintained, and shifted through the interaction. This makes AI system use a fundamentally social interaction. To build systems that are truly useful, we must define context as that which emerges in the dynamic interaction between AI system and user, and build an AI future which can adapt to the needs of people.

## Acknowledgments

The authors wish to thank the anonymous interview participants for their time and thoughts.

## References

[1] Syed AbuMusab. 2024. Generative AI and human labor: who is replaceable? *AI & SOCIETY* 39, 6 (2024), 3051–3053.

[2] Mark S. Ackerman. 2000. The Intellectual Challenge of CSCW: The Gap Between Social Requirements and Technical Feasibility. *Human–Computer Interaction* 15, 2 (2000), 179–203. doi:10.1207/S15327051HCI1523_5

[3] Philip E. Agre. 1997. Toward a Critical Technical Practice. In *Social Science, Technical Systems, and Cooperative Work: Beyond the Great Divide,* Geoffrey C. Bowker, Les Gasser, Susan Leigh Star, and Bill Turner (Eds.). Psychology Press, 131–158.

[4] Philip E Agre and David Chapman. 1987. Pengi: An implementation of a theory of activity. In *Proceedings of the sixth National conference on Artificial intelligence-Volume 1.* 268–272.

[5] Varol Akman. 2000. Rethinking context as a social construct. *Journal of pragmatics* 32, 6 (2000), 743–759.

[6] Humaid Al Naqbi, Zied Bahroun, and Vian Ahmed. 2024. Enhancing Work Productivity through Generative Artificial Intelligence: A Comprehensive Literature Review. *Sustainability* 16, 3 (2024). doi:10.3390/su16031166

[7] Humaid Al Naqbi, Zied Bahroun, and Vian Ahmed. 2024. Enhancing work productivity through generative artificial intelligence: A comprehensive literature review. *Sustainability* 16, 3 (2024), 1166.

[8] Abhijit Anand, Vinay Setty, Avishek Anand, et al. 2023. Context aware query rewriting for text rankers using llm. *arXiv preprint arXiv:2308.16753* (2023).

[9] Anthropic. 2025. Bringing Memory to Claude. *Anthropic Blog* (2025). https://www.anthropic.com/news/memory

[10] Suriya Ganesh Ayyamperumal and Limin Ge. 2024. Current state of LLM Risks and AI Guardrails. *arXiv preprint arXiv:2406.12934* (2024).

[11] Stefan Baack. 2024. A critical analysis of the largest source for generative ai training data: Common crawl. In *Proceedings of the 2024 ACM Conference on Fairness, Accountability, and Transparency.* 2199–2208.

[12] Yuntao Bai, Andy Jones, Kamal Ndousse, Amanda Askell, Anna Chen, Nova DasSarma, Dawn Drain, Stanislav Fort, Deep Ganguli, Tom Henighan, et al. 2022. Training a helpful and harmless assistant with reinforcement learning from human feedback. *arXiv preprint arXiv:2204.05862* (2022).

[13] Yushi Bai, Xin Lv, Jiajie Zhang, Yuze He, Ji Qi, Lei Hou, Jie Tang, Yuxiao Dong, and Juanzi Li. 2024. Longalign: A recipe for long context alignment of large language models. *arXiv preprint arXiv:2401.18058* (2024).

[14] Brian Balfour. 2025. The Next Great Distribution Shift. https://blog.brianbalfour.com/p/the-next-great-distribution-shift

[15] Ajay Bandi, Pydi Venkata Satya Ramesh Adapa, and Yudu Eswar Vinay Pratap Kumar Kuchi. 2023. The power of generative ai: A review of requirements, models, input–output formats, evaluation metrics, and challenges. *Future Internet* 15, 8 (2023), 260.

[16] Veronica Barassi. 2024. Toward a theory of AI errors: Making sense of hallucinations, catastrophic failures, and the fallacy of generative AI. *Harvard Data Science Review* Special Issue 5 (2024).

[17] Linda Liska Belgrave and Kapriskie Seide. 2019. Coding for grounded theory. *The SAGE handbook of current developments in grounded theory* (2019), 167–185.

[18] Stuart Bender. 2025. Generative-AI, the media industries, and the disappearance of human creative labour. *Media Practice and Education* 26, 2 (2025), 200–217.

[19] Debraj Bhattacharya. 2025. Context Is the New Moat in the AI-Driven Enterprise. *Medium* (2025). https://medium.com/@debraj.bhattacharya83/context-is-the-new-moat-in-the-ai-driven-enterprise-3d94d195e77b

[20] Paula Bialski. 2024. *Middle Tech: Software Work and the Culture of Good Enough.* Princeton University Press.

[21] Alexander Bick, Adam Blandin, and David Deming. 2025. *The Impact of Generative AI on Work Productivity.* Technical Report. Federal Reserve Bank of St. Louis. https://www.stlouisfed.org/on-the-economy/2025/feb/impact-generative-ai-work-productivity

[22] Andrea J Bingham and Patricia Witkowsky. 2021. Deductive and inductive approaches to qualitative data analysis. *Analyzing and interpreting qualitative data: After the interview* 1 (2021), 133–146.

[23] Ann E Blandford. 2013. *Semi-structured qualitative studies.* Interaction Design Foundation.

[24] Petter Bae Brandtzaeg and Marika Lüders. 2018. Time collapse in social media: Extending the context collapse. *Social Media+ Society* 4, 1 (2018), 2056305118763349.

[25] Kristin A Briney. 2024. Measuring data rot: An analysis of the continued availability of shared data from a Single University. *PloS one* 19, 6 (2024), e0304781.

[26] Tom Brown, Benjamin Mann, Nick Ryder, Melanie Subbiah, Jared D Kaplan, Prafulla Dhariwal, Arvind Neelakantan, Pranav Shyam, Girish Sastry, Amanda Askell, et al. 2020. Language models are few-shot learners. *Advances in neural information processing systems* 33 (2020), 1877–1901.

[27] Erik Brynjolfsson, Danielle Li, and Lindsey Raymond. 2025. Generative AI at work. *The Quarterly Journal of Economics* 140, 2 (2025), 889–942.

[28] Inha Cha and Richmond Y Wong. 2025. Understanding Socio-technical Factors Configuring AI Non-Use in UX Work Practices. In *Proceedings of the 2025 CHI Conference on Human Factors in Computing Systems.* 1–17.

[29] Aditya Challapally, Chris Pease, Ramesh Raskar, and Pradyumna Chari. 2025. The genai divide: State of ai in business 2025.

[30] Kaiyan Chang, Kun Wang, Nan Yang, Ying Wang, Dantong Jin, Wenlong Zhu, Zhirong Chen, Cangyuan Li, Hao Yan, Yunhao Zhou, et al. 2024. Data is all

you need: Finetuning llms for chip design via an automated design-data augmentation framework. In *Proceedings of the 61st ACM/IEEE Design Automation Conference*. 1–6.

[31] Yupeng Chang, Xu Wang, Jindong Wang, Yuan Wu, Linyi Yang, Kaijie Zhu, Hao Chen, Xiaoyuan Yi, Cunxiang Wang, Yidong Wang, et al. 2024. A survey on evaluation of large language models. *ACM transactions on intelligent systems and technology* 15, 3 (2024), 1–45.

[32] Aaron Chatterji, Thomas Cunningham, David Deming, Zoe Hitzig, Christopher Ong, Carl Yan Shan, and Kevin Wadman. 2025. *How People Use ChatGPT*. Technical Report w34255. National Bureau of Economic Research. w34255 pages. doi:10.3386/w34255

[33] Shreyas Chaudhari, Pranjal Aggarwal, Vishvak Murahari, Tanmay Rajpurohit, Ashwin Kalyan, Karthik Narasimhan, Ameet Deshpande, and Bruno Castro da Silva. 2025. Rlhf deciphered: A critical analysis of reinforcement learning from human feedback for llms. *Comput. Surveys* 58, 2 (2025), 1–37.

[34] Peter Baile Chen, Tomer Wolfson, Michael Cafarella, and Dan Roth. 2025. EnrichIndex: Using LLMs to Enrich Retrieval Indices Offline. *arXiv preprint arXiv:2504.03598* (2025).

[35] Paola Cillo and Gaia Rubera. 2025. Generative AI in innovation and marketing processes: A roadmap of research opportunities. *Journal of the Academy of Marketing Science* 53, 3 (2025), 684–701.

[36] A Feder Cooper, Emanuel Moss, Benjamin Laufer, and Helen Nissenbaum. 2022. Accountability in an algorithmic society: relationality, responsibility, and robustness in machine learning. In *Proceedings of the 2022 ACM conference on fairness, accountability, and transparency*. 864–876.

[37] Michael A Cusumano. 2023. Generative AI as a new innovation platform. *Commun. ACM* 66, 10 (2023), 18–21.

[38] Eve Danziger. 2010. Deixis, gesture, and cognition in spatial frame of reference typology. *Studies in Language. International Journal sponsored by the Foundation "Foundations of Language"* 34, 1 (2010), 167–185.

[39] Jenny L Davis and Nathan Jurgenson. 2014. Context collapse: Theorizing context collusions and collisions. *Information, communication & society* 17, 4 (2014), 476–485.

[40] Nassim Dehouche. 2021. Plagiarism in the age of massive Generative Pretrained Transformers (GPT-3). *Ethics in Science and Environmental Politics* 21 (2021), 17–23.

[41] Sarah Desrochers, James Wilson, and Matthew Beauchesne. 2024. Reducing hallucinations in large language models through contextual position encoding. (2024).

[42] Yi Dong, Ronghui Mu, Yanghao Zhang, Siqi Sun, Tianle Zhang, Changshun Wu, Gaojie Jin, Yi Qi, Jinwei Hu, Jie Meng, et al. 2025. Safeguarding large language models: A survey. *Artificial Intelligence Review* 58, 12 (2025), 382.

[43] Paul Dourish. 2004. What we talk about when we talk about context. *Personal and ubiquitous computing* 8, 1 (2004), 19–30.

[44] Amanda Dsouza, Christopher Glaze, Changho Shin, and Frederic Sala. 2024. Evaluating language model context windows: A" working memory" test and inference-time correction. *arXiv preprint arXiv:2407.03651* (2024).

[45] Vincent Enoasmo, Cedric Featherstonehaugh, Xavier Konstantinopoulos, and Zacharias Huntington. 2025. Structural embedding projection for contextual large language model inference. *arXiv preprint arXiv:2501.18826* (2025).

[46] Jad Esber, Sean Thielen-Esparza, Yondon Fu, and Tina He. 2025. Context is All You Need. *Available at SSRN 5277474* (2025).

[47] Wendy Nelson Espeland and Michael Sauder. 2007. Rankings and reactivity: How public measures recreate social worlds. *American journal of sociology* 113, 1 (2007), 1–40.

[48] I. John Fahrner, Emma Chen, Eric Topol, and Pranav Rajpurkar. 2025. The generative era of medical AI. *Cell* 188, 14 (2025), 3648–3660.

[49] Katja Fleischmann. 2024. Generative Artificial Intelligence in Graphic Design Education: A Student Perspective. *Canadian Journal of Learning and Technology* 50, 1 (2024), 1–17.

[50] Monika Fludernik. 1991. *Shifters and deixis: Some reflections on Jakobson, Jespersen, and reference.* Walter de Gruyter, Berlin/New York Berlin, New York.

[51] Ben Gansky and Sean McDonald. 2022. CounterFAccTual: How FAccT undermines its organizing principles. In *Proceedings of the 2022 ACM Conference on Fairness, Accountability, and Transparency*. 1982–1992.

[52] Yunfan Gao, Yun Xiong, Xinyu Gao, Kangxiang Jia, Jinliu Pan, Yuxi Bi, Yixin Dai, Jiawei Sun, Haofen Wang, and Haofen Wang. 2023. Retrieval-augmented generation for large language models: A survey. *arXiv preprint arXiv:2312.10997* 2, 1 (2023).

[53] Michail Giannakos, Roger Azevedo, Peter Brusilovsky, Mutlu Cukurova, Yannis Dimitriadis, Davinia Hernandez-Leo, Sanna Järvelä, Manolis Mavrikis, and Bart Rienties. 2025. The promise and challenges of generative AI in education. *Behaviour & Information Technology* 44, 11 (2025), 2518–2544.

[54] Barney G Glaser and Anselm L Strauss. 1998. Grounded theory. *Strategien qualitativer Forschung. Bern: Huber* 4 (1998).

[55] Dhruv Grewal, Cinthia B Satornino, Thomas Davenport, and Abhijit Guha. 2025. How generative AI is shaping the future of marketing. *Journal of the Academy of Marketing Science* 53, 3 (2025), 702–722.

[56] C. J. Gustafson. 2025. Context as a Competitive Moat. (2025). https://www.lookingforleverage.com/p/context-as-a-competitive-moat

[57] Manuel Hoffmann, Sam Boysel, Frank Nagle, Sida Peng, and Kevin Xu. 2024. *Generative AI and the Nature of Work.* Technical Report. CESifo Working Paper.

[58] Tarik Houichime and Younes El Amrani. 2025. Context Is All You Need: A Hybrid Attention-Based Method for Detecting Code Design Patterns. *IEEE Access* (2025).

[59] Lei Huang, Weijiang Yu, Weitao Ma, Weihong Zhong, Zhangyin Feng, Haotian Wang, Qianglong Chen, Weihua Peng, Xiaocheng Feng, Bing Qin, et al. 2025. A survey on hallucination in large language models: Principles, taxonomy, challenges, and open questions. *ACM Transactions on Information Systems* 43, 2 (2025), 1–55.

[60] Yuheng Huang, Jiayang Song, Zhijie Wang, Shengming Zhao, Huaming Chen, Felix Juefei-Xu, and Lei Ma. 2023. Look before you leap: An exploratory study of uncertainty measurement for large language models. *arXiv preprint arXiv:2307.10236* (2023).

[61] Xiang Hui, Oren Reshef, and Luofeng Zhou. 2024. The short-term effects of generative artificial intelligence on employment: Evidence from an online labor market. *Organization Science* 35, 6 (2024), 1977–1989.

[62] Dell Hymes. 2020. The scope of sociolinguistics. *International Journal of the Sociology of Language* 2020, 263 (2020), 67–76.

[63] Nanna Inie, Jeanette Falk, and Steve Tanimoto. 2023. Designing participatory ai: Creative professionals' worries and expectations about generative ai. In *Extended Abstracts of the 2023 CHI Conference on Human Factors in Computing Systems*. 1–8.

[64] Sunnie SY Kim, Elizabeth Anne Watkins, Olga Russakovsky, Ruth Fong, and Andrés Monroy-Hernández. 2023. Humans, ai, and context: Understanding end-users' trust in a real-world computer vision application. In *Proceedings of the 2023 ACM Conference on Fairness, Accountability, and Transparency*. 77–88.

[65] Heiko Koziolek, Sten Grüner, Rhaban Hark, Virendra Ashiwal, Sofia Linsbauer, and Nafise Eskandani. 2024. LLM-based and retrieval-augmented control code generation. In *Proceedings of the 1st International Workshop on Large Language Models for Code*. 22–29.

[66] Philippe Laban, Alexander R Fabbri, Caiming Xiong, and Chien-Sheng Wu. 2024. Summary of a haystack: A challenge to long-context llms and rag systems. *arXiv preprint arXiv:2407.01370* (2024).

[67] Oran Lang, Doron Yaya-Stupp, Ilana Traynis, Heather Cole-Lewis, Chloe R Bennett, Courtney R Lyles, Charles Lau, Michal Irani, Christopher Semturs, Dale R Webster, et al. 2024. Using generative AI to investigate medical imagery models and datasets. *EBioMedicine* 102 (2024).

[68] Matthew Law and Rama Adithya Varanasi. 2025. Generative AI and Changing Work: Systematic Review of Practitioner-Led Work Transformations Through the Lens of Job Crafting. In *International Conference on Human-Computer Interaction*. Springer, 131–152.

[69] Patrick Lewis, Ethan Perez, Aleksandra Piktus, Fabio Petroni, Vladimir Karpukhin, Naman Goyal, Heinrich Küttler, Mike Lewis, Wen-tau Yih, Tim Rocktäschel, et al. 2020. Retrieval-augmented generation for knowledge-intensive nlp tasks. *Advances in neural information processing systems* 33 (2020), 9459–9474.

[70] Belinda Z Li, Been Kim, and Zi Wang. 2025. QuestBench: Can LLMs ask the right question to acquire information in reasoning tasks? *arXiv preprint arXiv:2503.22674* (2025).

[71] Jiachang Liu, Dinghan Shen, Yizhe Zhang, Bill Dolan, Lawrence Carin, and Weizhu Chen. 2021. What Makes Good In-Context Examples for GPT-3? *arXiv preprint arXiv:2101.06804* (2021).

[72] Nelson F Liu, Kevin Lin, John Hewitt, Ashwin Paranjape, Michele Bevilacqua, Fabio Petroni, and Percy Liang. 2024. Lost in the middle: How language models use long contexts. *Transactions of the Association for Computational Linguistics* 12 (2024), 157–173.

[73] Xiaoou Liu, Tiejin Chen, Longchao Da, Chacha Chen, Zhen Lin, and Hua Wei. 2025. Uncertainty quantification and confidence calibration in large language models: A survey. In *Proceedings of the 31st ACM SIGKDD Conference on Knowledge Discovery and Data Mining V. 2*. 6107–6117.

[74] Sasha Luccioni, Yacine Jernite, and Emma Strubell. 2024. Power Hungry Processing: Watts Driving the Cost of AI Deployment?. In *Proceedings of the 2024 ACM Conference on Fairness, Accountability, and Transparency* (Rio de Janeiro, Brazil) *(FAccT '24)*. Association for Computing Machinery, New York, NY, USA, 85–99. doi:10.1145/3630106.3658542

[75] Donald MacKenzie. 2008. *An engine, not a camera: How financial models shape markets.* Mit Press.

[76] Michael Madaio, Lisa Egede, Hariharan Subramonyam, Jennifer Wortman Vaughan, and Hanna Wallach. 2022. Assessing the fairness of ai systems: Ai practitioners' processes, challenges, and needs for support. *Proceedings of the ACM on Human-Computer Interaction* 6, CSCW1 (2022), 1–26.

[77] Ramesh Manuvinakurike, Emanuel Moss, Elizabeth Anne Watkins, Saurav Sahay, Giuseppe Raffa, and Lama Nachman. 2025. Thoughts without Thinking: Reconsidering the Explanatory Value of Chain-of-Thought Reasoning in LLMs through Agentic Pipelines. *arXiv preprint arXiv:2505.00875* (2025).

[78] Alexander PL Martindale, Carrie D Llewellyn, Richard O De Visser, Benjamin Ng, Victoria Ngai, Aditya U Kale, Lavinia Ferrante di Ruffano, Robert M Golub, Gary S Collins, David Moher, et al. 2024. Concordance of randomised controlled trials for artificial intelligence interventions with the CONSORT-AI reporting guidelines. *Nature communications* 15, 1 (2024), 1619.

[79] Alice E Marwick and danah boyd. 2011. I tweet honestly, I tweet passionately: Twitter users, context collapse, and the imagined audience. *New media & society* 13, 1 (2011), 114–133.

[80] Timothy R McIntosh, Teo Susnjak, Nalin Arachchilage, Tong Liu, Dan Xu, Paul Watters, and Malka N Halgamuge. 2025. Inadequacies of large language model benchmarks in the era of generative artificial intelligence. *IEEE Transactions on Artificial Intelligence* (2025).

[81] Albert Mehrabian et al. 1971. *Silent messages.* Vol. 8. Wadsworth Belmont, CA.

[82] Lingrui Mei, Jiayu Yao, Yuyao Ge, Yiwei Wang, Baolong Bi, Yujun Cai, Jiazhi Liu, Mingyu Li, Zhong-Zhi Li, Duzhen Zhang, et al. 2025. A Survey of Context Engineering for Large Language Models. *arXiv preprint arXiv:2507.13334* (2025).

[83] Celestine Mendler-Dünner, Gabriele Carovano, and Moritz Hardt. 2024. An engine not a camera: Measuring performative power of online search. *Advances in Neural Information Processing Systems* 37 (2024), 59266–59288.

[84] Uday Mittal, Siva Sai, Vinay Chamola, and Devika Sangwan. 2024. A comprehensive review on generative AI for education. *Ieee Access* 12 (2024), 142733–142759.

[85] Ali Modarressi, Hanieh Deilamsalehy, Franck Dernoncourt, Trung Bui, Ryan A. Rossi, Seunghyun Yoon, and Hinrich Schütze. 2025. NoLiMa: Long-Context Evaluation Beyond Literal Matching. arXiv:2502.05167 [cs.CL] https://arxiv.org/abs/2502.05167

[86] Emanuel Moss. 2025. Toward a Realpolitik for AI. *Public Books* (2025). https://www.publicbooks.org/toward-a-realpolitik-for-ai/

[87] Emanuel Moss, Elizabeth Watkins, Christopher Persaud, Passant Karunaratne, and Dawn Nafus. 2025. Controlling Context: Generative AI at Work in Integrated Circuit Design and Other High-Precision Domains. *arXiv preprint arXiv:2506.14567* (2025).

[88] Subigya Nepal, Javier Hernandez, Talie Massachi, Kael Rowan, Judith Amores, Jina Suh, Gonzalo Ramos, Brian Houck, Shamsi T Iqbal, and Mary P Czerwinski. 2025. From User Surveys to Telemetry-Driven AI Agents: Exploring the Potential of Personalized Productivity Solutions. *Proceedings of the ACM on Human-Computer Interaction* 9, 7 (2025), 1–41.

[89] Cal Newport. 2025. Why can't A.I. manage my e-mails? https://www.newyorker.com/culture/open-questions/why-cant-ai-manage-my-e-mails

[90] Davide Nicolini. 2012. *Practice theory, work, and organization: An introduction.* OUP Oxford.

[91] Kate Niderhoffer, Gabriella Rosen Kellerman, Angela Lee, Alex Liebscher, Kristina Rapuano, and Jeffrey T. Hancock. 2025. AI-Generated "Workslop" Is Destroying Productivity. *Harvard Business Review* (2025). https://hbr.org/2025/09/ai-generated-workslop-is-destroying-productivity

[92] Helen Nissenbaum. 2010. *Privacy in context: technology, policy, and the integrity of social life.* Stanford Law Books.

[93] Helen Nissenbaum. 2004. Privacy as contextual integrity. *Wash. L. Rev.* 79 (2004), 119.

[94] Vidushi Ojha, Andrea Watkins, Christopher Perdriau, Kathleen Isenegger, and Colleen M Lewis. 2024. Instructional Transparency: Just to Be Clear, It's a Good Thing. In *Proceedings of the 2024 ACM Conference on International Computing Education Research-Volume 1.* 192–205.

[95] Srishti Palani and Gonzalo Ramos. 2024. Evolving roles and workflows of creative practitioners in the age of generative AI. In *Proceedings of the 16th Conference on Creativity & Cognition.* 170–184.

[96] Carl Preiksaitis and Christian Rose. 2023. Opportunities, challenges, and future directions of generative artificial intelligence in medical education: scoping review. *JMIR medical education* 9 (2023), e48785.

[97] Ting Qiu, Di Yang, Hui Zeng, and Xinghao Chen. 2024. Understanding graphic designers' usage behavior of generative artificial intelligence tools. *Kybernetes* (2024).

[98] Anthony Rhodes, Ramesh Manuvinakurike, Sovan Biswas, Giuseppe Raffa, and Lama Nachman. 2025. Uncertainty Quantification with Generative-Semantic Entropy Estimation for Large Language Models. https://openreview.net/forum?id=LDmJFJlo83

[99] Daniel Russo. 2024. Navigating the complexity of generative ai adoption in software engineering. *ACM Transactions on Software Engineering and Methodology* 33, 5 (2024), 1–50.

[100] Gery W Ryan and H Russell Bernard. 2003. Techniques to identify themes. *Field methods* 15, 1 (2003), 85–109.

[101] Daniel Schiff, Aladdin Ayesh, Laura Musikanski, and John C Havens. 2020. IEEE 7010: A new standard for assessing the well-being implications of artificial intelligence. In *2020 IEEE international conference on systems, man, and cybernetics (SMC).* IEEE, 2746–2753.

[102] Deborah Schiffrin. 2005. Discourse markers: Language, meaning, and context. *The handbook of discourse analysis* (2005), 54–75.

[103] Philipp Schmid. 2025. *The New Skill in AI is Not Prompting, It's Context Engineering.* https://www.philschmid.de/context-engineering

[104] Reva Schwartz, Rumman Chowdhury, Akash Kundu, Heather Frase, Marzieh Fadaee, Tom David, Gabriella Waters, Afaf Taik, Morgan Briggs, Patrick Hall, Shomik Jain, Kyra Yee, Spencer Thomas, Sundeep Bhandari, Paul Duncan, Andrew Thompson, Maya Carlyle, Qinghua Lu, Matthew Holmes, and Theodora Skeadas. 2025. Reality Check: A New Evaluation Ecosystem Is Necessary to Understand AI's Real World Effects. arXiv:2505.18893 [cs.CY] https://arxiv.org/abs/2505.18893

[105] Nick Seaver. 2015. The nice thing about context is that everyone has it. *Media, Culture & Society* 37, 7 (2015), 1101–1109.

[106] Mrinank Sharma, Meg Tong, Tomasz Korbak, David Duvenaud, Amanda Askell, Samuel R. Bowman, Esin DURMUS, Zac Hatfield-Dodds, Scott R Johnston, Shauna M Kravec, Timothy Maxwell, Sam McCandlish, Kamal Ndousse, Oliver Rausch, Nicholas Schiefer, Da Yan, Miranda Zhang, and Ethan Perez. 2024. Towards Understanding Sycophancy in Language Models. In *The Twelfth International Conference on Learning Representations.* https://openreview.net/forum?id=tvhaxkMKAn

[107] Katie Shilton, Emanuel Moss, Sarah A Gilbert, Matthew J Bietz, Casey Fiesler, Jacob Metcalf, Jessica Vitak, and Michael Zimmer. 2021. Excavating awareness and power in data science: A manifesto for trustworthy pervasive data research. *Big Data & Society* 8, 2 (2021), 20539517211040759.

[108] Mona Sloane, David Danks, and Emanuel Moss. 2024. Tackling ai hyping. *AI and Ethics* 4, 3 (2024), 669–677.

[109] Mona Sloane, Ian Rene Solano-Kamaiko, Jun Yuan, Aritra Dasgupta, and Julia Stoyanovich. 2023. Introducing contextual transparency for automated decision systems. *Nature Machine Intelligence* 5, 3 (2023), 187–195.

[110] Vicki Smith. 1997. New forms of work organization. *Annual review of sociology* 23, 1 (1997), 315–339.

[111] Beth Stackpole. 2024. The impact of generative AI as a general-purpose technology. *MIT Sloan Management Review* (2024). https://mitsloan.mit.edu/ideas-made-to-matter/impact-generative-ai-a-general-purpose-technology

[112] Yuan Sun, Eunchae Jang, Fenglong Ma, and Ting Wang. 2024. Generative AI in the wild: Prospects, challenges, and strategies. In *Proceedings of the 2024 CHI Conference on Human Factors in Computing Systems.* 1–16.

[113] Yujie Sun, Dongfang Sheng, Zihan Zhou, and Yifei Wu. 2024. AI hallucination: towards a comprehensive classification of distorted information in artificial intelligence-generated content. *Humanities and Social Sciences Communications* 11, 1 (2024), 1–14.

[114] Mirac Suzgun and Adam Tauman Kalai. 2024. Meta-prompting: Enhancing language models with task-agnostic scaffolding. *arXiv preprint arXiv:2401.12954* (2024).

[115] Ashish Vaswani, Noam Shazeer, Niki Parmar, Jakob Uszkoreit, Llion Jones, Aidan N Gomez, Łukasz Kaiser, and Illia Polosukhin. 2017. Attention is all you need. *Advances in neural information processing systems* 30 (2017).

[116] Pablo Villalobos, Anson Ho, Jaime Sevilla, Tamay Besiroglu, Lennart Heim, and Marius Hobbhahn. 2024. Position: Will we run out of data? Limits of LLM scaling based on human-generated data. In *Forty-first International Conference on Machine Learning.*

[117] Jessica Vitak. 2012. The impact of context collapse and privacy on social network site disclosures. *Journal of broadcasting & electronic media* 56, 4 (2012), 451–470.

[118] Tung Vuong, Salvatore Andolina, Giulio Jacucci, and Tuukka Ruotsalo. 2021. Does more context help? Effects of context window and application source on retrieval performance. *ACM Transactions on Information Systems (TOIS)* 40, 2 (2021), 1–40.

[119] Xinyuan Wang, Liang Wu, Liangjie Hong, Hao Liu, and Yanjie Fu. 2025. LLM-Enhanced User–Item Interactions: Leveraging Edge Information for Optimized Recommendations. *ACM Transactions on Intelligent Systems and Technology* 16, 5 (2025), 1–24.

[120] Elizabeth Anne Watkins, Emanuel Moss, Giuseppe Raffa, and Lama Nachman. 2025. What's So Human about Human-AI Collaboration, Anyway? Generative AI and Human-Computer Interaction. arXiv:2503.05926 [cs.HC] https://arxiv.org/abs/2503.05926

[121] Geoffrey I Webb, Roy Hyde, Hong Cao, Hai Long Nguyen, and Francois Petitjean. 2016. Characterizing concept drift. *Data Mining and Knowledge Discovery* 30, 4 (2016), 964–994.

[122] Alina Wernick, Alan Medlar, Sofia Söderholm, and Dorota Głowacka. 2025. Evaluating the Contextual Integrity of False Positives in Algorithmic Travel Surveillance. In *Proceedings of the 2025 ACM Conference on Fairness, Accountability, and Transparency.* 2056–2070.

[123] Michael Wessel, Martin Adam, Alexander Benlian, Ann Majchrzak, and Ferdinand Thies. 2025. Generative AI and its transformative value for digital platforms. *Journal of Management Information Systems* 42, 2 (2025), 346–369.

[124] Amy Winograd. 2022. Loose-lipped large language models spill your secrets: The privacy implications of large language models. *Harv. JL & Tech.* 36 (2022), 615.

[125] Allison Woodruff, Renee Shelby, Patrick Gage Kelley, Steven Rousso-Schindler, Jamila Smith-Loud, and Lauren Wilcox. 2024. How knowledge workers think generative ai will (not) transform their industries. In *Proceedings of the 2024 CHI Conference on Human Factors in Computing Systems.* 1–26.

[126] Yuhao Wu, Evin Jaff, Ke Yang, Ning Zhang, and Umar Iqbal. 2025. An in-depth investigation of data collection in llm app ecosystems. In *Proceedings of the 2025 ACM Internet Measurement Conference*. 150–170.

[127] Xiaodan Xing, Fadong Shi, Jiahao Huang, Yinzhe Wu, Yang Nan, Sheng Zhang, Yingying Fang, Michael Roberts, Carola-Bibiane Schönlieb, Javier Del Ser, and Guang Yang. 2025. On the caveats of AI autophagy. *Nature Machine Intelligence* 7, 2 (Feb. 2025), 172–180. doi:10.1038/s42256-025-00984-1

[128] Peixi Xiong, Chaunte W. Lacewell, Sameh Gobriel, and Nilesh Jain. 2025. Context Is All You Need: Efficient Retrieval Augmented Generation for Domain Specific AI. In *First Workshop on Scalable Optimization for Efficient and Adaptive Foundation Models*. https://openreview.net/forum?id=6bKHoUQWlo

[129] Peng Zhang and Maged N Kamel Boulos. 2023. Generative AI in medicine and healthcare: promises, opportunities and challenges. *Future Internet* 15, 9 (2023), 286.

[130] Wayne Xin Zhao, Kun Zhou, Junyi Li, Tianyi Tang, Xiaolei Wang, Yupeng Hou, Yingqian Min, Beichen Zhang, Junjie Zhang, Zican Dong, et al. 2023. A survey of large language models. *arXiv preprint arXiv:2303.18223* 1, 2 (2023).

[131] Michael Zimmer. 2018. Addressing conceptual gaps in big data research ethics: An application of contextual integrity. *Social Media+ Society* 4, 2 (2018), 2056305118768300.

## A Generative AI Usage Statement

Generative AI was not used to draft any portion of this paper's text. Interviews were transcribed using speech-to-text tools that embed generative AI. Every effort was taken to verify quotes reproduced above using the original recording. Transcripts were analyzed using Dovetail, which embeds generative AI tools in an intransparent matter, however summaries auto-generated in Dovetail were purposefully ignored by analysts. Generative AI tools were used for producing cursory briefs on existing literature during the ideation phase of the paper, but were not incorporated into the final text in any way.

## B Appendix

### B.1 Prompts

| | | |
|---|---|---|
| Prompt 01 | P01 | So if it gave me something that was kind of in the realms I'm looking for, maybe stylistically I wasn't sure of which way to go. And I'd say give me some ideas that would incorporate more of a vintage illustration design. And then it would list off a couple things. And then I'd be like, well, can we focus more on like the art deco kind of aspect of that? And it would just kind of progress like that. Cool. That makes a ton of sense. I'm curious. Like, you're describing very aesthetic features, but it sounds like it's a text based interaction. Like what kind of things can it suggest to make something more art deco or more vintage? Just to inform. Inform me, I guess, more of how the design is put together visually. Like with art deco and like the 60s, 70s, you got a lot of geometry, shapes and things like that. So those would be the kind of things that it would use. It would suggest using maybe more geometric layouts or composition Gotcha. So it actually knows the conventions of Art Deco pretty well from like a linguistic. Like being able to describe it in language. |
| Prompt 02 | P01 | My clients starting to do auctions on whatnot this week. My first show or their first shows this week. Company's name is Midnight Collectibles. Could you help me brainstorm a title for the auctions and other ideas for branding and marketing? |
| Prompt 03 | P01 | I'll say, can you please assist me with brainstorming some creative ideas for this apparel design I want to do? They want it to include some esoteric aspects along with astrology, and they also want to incorporate these colors with it. What are some ways I could go about approaching this? Cool. Gotcha. And then does it need usually more nudging or more context? It almost turns into like a back and forth, like, oh, give me some different ideas and I'll say, well, what if we tried this with that one? And then I'll just kind of progress. Almost like I'm working with an assistant on trying to brainstorm and break things down. |
| Prompt 04 | P01 | please assist me with writing a proposal for this job posting. Please include my experience, my current work flow, and. And then anything else that would be included that would kind of correlate with it. And then I would also put. For some reason this seems to help. If I grab what the actual job listing is and put it on there afterwards in parentheses, it seems to help it write it out better. I don't know why that's interesting. |
| Prompt 05 | P02 | I have another prompt that I asked it to create a search string for me based on. I asked it to create three search strings for me based on the concepts in the article. And then those search strings are structured so that they're searching Bloomberg, Al Jazeera, Reuters, and ap. And so if I found something that I'm just not getting the results I want, I get it to create a search string for me. I copy and paste that into Google. If I find a better written version of the same story, I will run the query again on that story and I usually get something that I can use at that point. |

| Prompt 06 | P02 | So I have a prompt. So what I primarily use Poe for is I use it to access Gemini in that interface because I find the interface comfortable and I like the way that Gemini phrases things. I have a prompt that I wrote, so I use it to summarize these news articles on the terrorism financing front. This is my. This is the prompt that I use. It's write an eight sentence summary of this article in the third person emphasizing financing, fundraising and earning money. Use terrorism instead of terror. Use us together, not u s. If foreign currency is mentioned, put the US dollar amount next to it in a specific format that we use. And then, yeah, that's the extent of my prompt. And so I find that eight sentences gives me enough information that I can edit it down to the two to three sentences that go in the newsletter for each of these individual stories. It usually gives me the names that I need. It gives me the amounts. It tends to format things appropriately with the, with the acronyms spelling out the acronyms. So I either will add or remove acronyms in order to sort of fit our conventions for the newsletter. And it just. It phrases it in the third person in a way. There are. There are artifacts that I go through and I edit out afterwards. A lot of the LLMs use utilize constantly. |
|---|---|---|
| Prompt 07 | P02 | one of the prompts that I use on a fairly regular basis is along the lines of break this article down into sections and summarize each section for me. When I type N, move on to the next section and so it'll scroll through the page and it'll summarize each section. And if I spot something that I need, I can just very quickly pop down to that section. I can read through it as a whole piece, but breaking the information down into bite sized chunks in the same sort of way that I teach students to read an academic article. |
| Prompt 08 | P02 | Yeah, so the first part of that was write me a five sentence structure of this article or five sentence summary of this article. And then I added in the emphasizing financing, fundraising and earning money. And that seems to get the topics I had it much more focused on like money laundering and terrorism financing. But that ended up being too Specific and it kept using the word terrorism financing. Terrorism financing. Terrorism financing. The formatting stuff is is sort of like some basics from our house style in the newsletter. So terrorism is always used instead of terror. US is formatted in a certain way. US Dollars. That used to take a lot of time, but having the LLM do the currency conversion has saved me a huge amount of time. So the section if foreign currency is mentioned, put the US dollar amount next to it like this. And then I just give an example of the formatting for the currency conversion. The third person is an actual. That's a recent addition and that was because it started producing some stuff occasionally in the second person, which was strange, but that was an easy thing to fix by just adding that element in there. The other one the search string one Here is Create three simple Google search strings on different topics to help find pages on similar topics. Use full names of people and organizations if there is a connection to the United states. Include site justice.gov always include site Al Jazeera cite Reuters, site AP news and site Bloomberg. Separate all site search links with or do not use date limits. Do not use quotation marks. Return only the search strings in a list with no explanation. |
| Prompt 09 | P03 | Okay. Change it to be a little bit more Midwestern. |
| Prompt 10 | P03 | I've had to change up the prompts a little bit. Ask it in a different way. You know, tell me like you. Like you were talking to a fifth grader, you know, that kind of stuff. But not so much in my writing. |
| Prompt 11 | P04 | I prompted it something about, you know, what to work out plan for like a 40 year old golfer in the off season. And so I was trying to figure out like okay, how can this kind of come up with a plan? And it then seemed to come back with well, you should do cardio two days a week, you should do strength train three times a week, you should do some stretching, you should do some golf specific exercises. It just seemed like it came back with a plan that, that might take, excuse me, it might take three hours a day if I do all of the things that it's telling me without it really kind of giving because it mentioned, you know, I could walking, cycling or swimming for 30 minutes. |
| Prompt 12 | P05 | I was telling it to put in long term memory about some of my favorite bands, which started making me think about Marduk funeral, Ms. Behemoth, Belphegor, Megala, etc. |
| Prompt 13 | P05 | Absolutely. So if I'm generating something for a patient I'm going to use a very specific prompt. I will quickly say, you know, a, you know, the 36 year old female in new diagnosis of you know, pre diabetes, please generate some patient friendly discharge instructions to this patient including resources to nutritionfacts.org for her and some whole food plant based healthy eating tips. Put this in paragraph format. Make this less than one page. Make this at the 8th grade level. Very specific things like that. |

| Prompt 14 | P05 | Oh, I completely forgot. We need to put this other piece of information in there. We're also struggling with lipids. So let's add, let's just add that. Then I'll just say, oh, also please add in some info about hyperlipidemia. Very accessible for the patient. And tie that in. Tie in. Why it's important to manage lipids in someone who's pre diabetic. We have patient handouts like, like 40,000 of them with our. In every language you can imagine. But they're so nonspecific that I don't like any of them. They're. They're so up here and they don't, they don't tailor anything. But this tool allows me to tailor the patient education every time. And it's so fast that it's doing the patient a much better service to give them this tailored stuff that takes seconds to do as opposed to just a sheet on lipids or whatever that our hospital has written. |
|---|---|---|
| Prompt 15 | P06 | I will start high level because I want to see what it gives me and then I refine from there. |
| Prompt 16 | P06 | quite often, you know, you have, you have states, they mirror each other's laws when they kind of exist next to each other. You know, you have a Louisiana and Alabama or something of that nature. They're like, oh, we've leveraged off each other's laws. And I was like, show me the comparisons. What's different? So if I'm working this project and considering these two states that, that might have very similar, you know, laws. Well, I can reduce my Work time knowing I can move the project, you know, in dual format in a way. So I'll do things like that. And also I'll revise my, my query based off. If I'm not getting information, |
| Prompt 17 | P06 | Yeah, so then it starts bringing me and, you know, if it doesn't get, give me a link. I'll also go, you know, or if it doesn't cite a statute, I go, what, what statute governs, you know, timeshare in the state of Texas? I will get into that. And then when I get there, you know, I might click into that resource to see what it's giving me. But then I might come back going, you know, bullet point for me, you know, so and so forth. Something I might also do, let's just say an example. In this scenario, I'm like, oh, I've got two states that have to search, you know, the same information. I'll ask it to compare the differences. |
| Prompt 18 | P07 | And I said that I just needed a more comprehensive snapshot. And I said, an example of what I needed is narrated in this document. And then I said, can you do it on these different regions? I want it to be more easy to read, more eye catching, keep it to one to two pages, export it to Word. And then I said, feel free to add extra findings or data that make the document more compelling. Like if it's looking at the data and it's like, oh, that's interesting, go ahead and pull that out. And keeping the focus around, like keeping children, their families who care about them and not putting them in foster care. And so then I put in some more information and it spit this out. And then it gave me something and I was like, well, what report was that in? Because it wasn't Matching up, like, what I was seeing. And it told me it was in this snapshot. And I was like, oh, okay. |
| Prompt 19 | P07 | Okay, here's what I did for my partner. This is a really interesting one. For Valentine's Day, I was having his car detailed inside and out, and I wanted to make him a card with, like, a riddle. And so it kind of like, gave me a riddle for him to start to guess. And I said, like, the second One, can you make it a little bit longer? And then it made it a little bit longer. And then I said, do you have an alternate version? I was just curious. And then they gave me an alternate version. And I must have went with that or the one before it. And it was just like, you know, I just wanted something silly because, like I wasn't going to be able to give him the car detail that day. |
| Prompt 20 | P08 | how do I converse with a client who needs. Whose needs are beyond my capabilities, but if given different parameters, something like that. I feel like, yeah, I remember seeing something about like the first bit. I definitely do already. And that's what I was saying. And then this is. This second bit is what I. What I was talking about is like the helpful bit most of the time being firm, being explained. And. And then this. Yeah, this one might have been a bit more open chat. I feel like there was something that I further chose somewhere down the line. It was the second part. I think I said, what if I already being topic and direct also. |

| Prompt 21 | P08 | I'm really intrigued by, like, all the progressive and, like, harmonious, humanistic, if you will, elements of AI. I feel like there's a lot of, like, gems just in, like, the simplicity of asking about something as simple as, like, health care or like, education or like, family dynamics or like, war and the world at large and that it just. It'll give you an answer that is, like, very pacifistic a lot of times, and like, a very, like, just like, all around beautiful answer. And so there's been. There's been conversations I've had where it just gives me this answer. Like, I think I was like, how do I stop. How do we stop war on Earth? And something. And then it gave you something back that was like, well, all the nations will need to work together and then, like, we'll have to have, like, more structure throughout everything. And then I just loved, like, I was. It's like I was, like, pulling on this, like, thread of, like, AI wisdom that was just coming out. And it's so simple if you think about it. It's like something like a child would respond with that doesn't understand, like, global tensions or something. It was just like, everyone needs to work together. So did you. |
|-----------|-----|----|
| Prompt 22 | P09 | Because a lot of times a drug like say Humira. Oh, well, you have to try the generic version of this, this, this, this before you get this medication. Well, the patient's not gonna understand that. So I could take that guideline denial post into copilot. I could ask like, hey, you know, I'm writing a. And then two. We sometimes we have to formulate our denials based off of the patient speaks English or Spanish. So if I need like that. Yeah, so I can translate. |
| Prompt 23 | P09 | So I could use something for like translation. But I could take that denial from the guidelines tab posting co pilot type of prompt. Like, okay, I am reviewing a prioritization that results in denial for a member. I'm meaning to include this specific rationale for denial in this but put it in a. Make it to where even like a fifth grader could read it. Something like that. Like, you know, like dumb down the, like be specific with the prompt because you have to be specific to get like a specific answer. But put in stuff like that that will be able to make it understandable for the patient for why it denied that type of stuff. |
| Prompt 24 | P10 | Okay. So I do see one thing on here at the top, it says, give me 10 options for clever business name selling Grandma, Nana, and Gigi gifts. So then, you know, it gave me a list. And I remember, Let me see. Putting in here, okay. And I put in here, I like these, but will have mom items too. So initially I had just said, you know. And so it gave me a list, you know, some options, 10 options. And then I know it's. I remember wanting a different word for matriarch, but with the same feel as a matriarch. And then I remember putting in there, I like this. What is it? The. There's one saying out there, it says, my favorite people call me mom and Gigi. And so I put in there, I like this saying, give me something similar to it with the same feel. And then I remember going back and saying, I like this word, but put it in a different way. And let's see, give me 10 options to show the bond, to indicate the bond between a grandchild and grandparent. So just different things like that. Because, you know, the different options that it was giving me, I liked, but just was trying to. Just trying to find the perfect thing. |
| Prompt 25 | P11 | For example, like before generating this video, you have to let Sora understand what hepatocytes are. So hepatocytes are basically like liver cells. So let ChatGPT generates a detailed content about explaining the morphologies of the hepatocytes. Right. And also tell. Let ChatGPT generates the functions of the cells and for example, most of the cells only have one nucleus, but this one has two nucleuses. So you have to specify this or you can make mistakes. And all the scientists understand the behind story. So they would point out this is not correct. So you have to be very careful about these contents. So let's say how that specific component migrates into the cells. So you have to tell like the pathway. So let's chatgpt generate the detailed pathway, how it feels the cell membrane, how it goes to the nucleus, how it edits the gene, how it's unwinding the DNA and stuff like that. So you have to be very, very detailed in order to generate the video. Because as I mentioned, generating video can be very expensive. So for generating images it's easy. Like you let CH generate paragraph first. Based on that, just generate image, you make further changes when it gets it. |
| Prompt 26 | P12 | Let's say I'm working on quantitation of pathogen in urine and to understand the scaling factors being used in converting the relative quantitation into absolute quantitation. I'm just simply going to ChatGPT, type in the prompt and just to understand what ChatGPT give me about the scaling factor and also give me a breakdown of how they did the conversion and how the step-by-step processes starting from sample prep and the method used. So I use that what ChatGPT has given me and compare with what I have. |

| Prompt 27 | P13 | Like if you have a prior study, you might say, 'well, this is new. In the last three months,' you might say, 'when I gave contrast, it enhanced, but only at the periphery.' You might ask it like, 'what other imaging should be done? If it's one of these two things, like, what else could you do to figure out what this is?' Because often the clinician wants to, okay, what kind of follow up do I have to do before I send them home? Like schedule for a PET scan, a biopsy, MRI, what are we doing? Because they don't know imaging and they need to know that stuff. |
|---|---|---|

**Table 10: Selected quotes from participants.**

## B.2 Quotes

| | | |
|---|---|---|
| Q01 | P02 | "For generative AI, I have NotebookLM. I have 40 books that are really influential to me that I have in NotebookLM and then I'll query against that. I also have my conversation history on a question and answer forum so that I can go back and I can like have chats with myself and, and look up things that I've written in the past in terms of the stuff that I have that I use on a regular basis." |
| Q02 | P05 | "... if you say to ChatGPT store in your long term memory, whatever, it will lock that into its long term memory. And so I have pasted a ton of stuff that I've written into it. And so I had to respond to this, you know, this like silly interview the other day for the psych department about like, oh, we're going to feature you on Instagram. Answer these questions. I did not have time to do this. And so it has read hundreds of pages of my writing and it stored it in long term memory. So I said just respond to these questions, knowing what you know about me. And it responded two sentences to each of them. It sounded exactly like me, you know, and it was an appropriate answer to all those questions. I made minimal changes." |
| Q03 | P05 | "So I use that tool [Doximity] all the time at work to generate documents. For instance, if I need to generate a hospital course summary, I can paste in maybe 20, 30 pages of information from someone's hospital course, their admission, HMP history and physical, their most recent progress note, most recent consultant notes, and then it will generate a timeline of exactly what happened while they were in the hospital, when this MRI came back, that kind of thing." |
| Q04 | P07 | "I use it for taking like a lot of my thoughts or like my random like notes and just like typing it all to chatgpt. And I say like make sense of this and put it into a document and it kind of gives me a baseline for a narrative that I typically usually have to kind of flesh out a little bit more, make it a little bit more specific or more detailed to the actual event or whatever I'm reporting on. But it's a really good way to just get... like I have all of these flip charts from this event, I have all these sticky notes from the event. They're all typed up. So like let's get it into like a readable format, like four pages maybe." |
| Q05 | P09 | "The system like the CCE tool can pull information, all these files for me that I need, save me time from having to look through all those documents. So it highlights information I need to be able to pull and complete the case. And we have a system that automatically generates what we call criteria. The criteria is the form that's used to get whatever drug is being requested, approved or denied. So there's a system that populates the questions based off the answers that are provided. And that's a system that's, you know, internally built by the company. I don't know what is called cast, but each specific insurance plans guidelines are in there, we have to follow that." |
| Q06 | P11 | "Q: Okay, so the company specific ChatGPT that you mentioned that [vendor] built, does it have access to proprietary information? Can you ask questions about your own products and get some answers? A: Yeah, I think so, yes. So the because it's company's products so we are free to type in whatever we want." |
| Q07 | P07 | "And I do try to keep keep themes together on my threads. I don't do like a new thread every time. Like I'll try to go back to like [a specific project] and kind of continue that conversation." |
| Q08 | P05 | "If I'm writing patient instructions, I want it usually at the fifth or eighth grade level. But if I'm writing a discharge summary hospital course that is not for patients... it should be written in formal technical language that is accessible to those other physicians that are reading this. So that's why I put that information in there." |
| Q09 | P09 | " I'm meaning to include this specific rationale for denial in this but ... make it to where even like a fifth grader could read it. Like ... be specific with the prompt because you have to be specific to get like a specific answer. But put in stuff like that that will be able to make it understandable for the patient." |
| Q10 | P07 | "It's like a family group chat. My mom started it, and he [my nephew] was really going against her. And... this was the one time I could actually ... defend my mom, which is ... rare. So I would like, copy, what he said on my phone and then put that into ChatGPT and say, how should I respond to this? And then I would copy it and put it back. And then, yeah, just kind of went back and forth." |
| Q11 | P02 | "I am looking for tools that can act in the same way that I act." |
| Q12 | P06 | "Get me out of a tourist trap, find me a local experience, please. I'm going to be very plain language, not tourist trap estaurants. Include unique, hole in the wall, places that exceed expectations." |
| Q13 | P10 | "Mostly I just wanted to see what was in walking distance from there. So that would kind of narrow my plans down of what we're going to do is something that would, you know, be close enough to walk." |
| Q14 | P05 | "But with Open Evidence it's just simply a search engine that gives you specific tailored evidence based results that pull from papers. And so it's basically just a faster way of searching PubMed using advanced search... The attending would ask ... my intern: With their CKD [chronic kidney disease], can we start Entresto? And and they would go, I don't know. And I'll go, listen you guys, move on to the next system, I'll look this up on Open Evidence. And then they finish talking about, you know, what we're going to do from a protein calorie malnutrition standpoint, which, say, another problem the patient has. And by the time that's done, it's been 30 seconds and I've pulled this up. Okay, you know what? I think it is fine to start that. Let's wait for today's echo to come back, and then we'll go from there. So that's how it would work in that flow." |
| Q15 | P05 | "I've trained my private one to be extremely, like, you know, like, loose with me and very blunt and honest and funny. And it will be a bit irreverent. And so it will be like, you know, "Yo, Captain [Lastname]!"." |

| Q16 | P02 | " I find that eight sentences gives me enough information that I can edit it down to the two to three sentences that go in the newsletter for each of these individual stories." |
|-----|-----|---|
| Q17 | P03 | "It needs to be helpful, useful, informative. I'm sure the customer reads it when he gets it and looks it over to see if there's any serious problems. They can ask for a revision, although nobody ever does rarely anymore. And they can go in, maybe change some things too if they want to." |
| Q18 | P06 | "I adjust it to my style. I always still tweak the final response. I take word processor or something. Yes, I will take it over to word, you know, because I always [personalize it]." |
| Q19 | P07 | "I love it for briefing documents. Master's programs are all reading and I don't have time to read all of it, so it'll give me really good overview for the most part, at least enough to get through class, which is exactly what I need. I've had it, like, put things in tables for me so I can see it in a different way." |
| Q20 | P04 | "Another one that doesn't have much back and forth is I've used it sometimes just to help teach me a topic. Sometimes it's teaching me a topic that I know, but I want to see how it tells me so that then maybe I can try teaching someone else in a different way. Like, like I understand this, but I need some better ways to include it in a presentation. So I've done that." |
| Q21 | P06 | "A lot of times I'll use it as a brainstorming tool. And I know that when I brainstorm, I'll say, rewrite the following paragraph in first person in the same tone in paragraph form and any other things." |
| Q22 | P13 | "But sometimes you just want ChatGPT or something you can put things in. I'm always hesitant 'cause. I'm like, should I trust it? But at least it's a starting point. Then I can say OK. This sounds reasonable. Let me go look that up. Look up some of the suggestions now in STATdx or something and see if that makes sense. And often that will have what's called a differential diagnosis, where, well, the adrenal adenoma.... here's other things it could be. Look for these findings and things, and so it gives you somewhere to start. If you're having trouble, like getting your feet on the ground to know like. OK, what is this? I've never seen this, never saw this in residency. Never seen a case like this. There's no history to tell me what. Strange sarcoma. Whatever tumor this guy has like. So sometimes it can help you like narrow things down, or at least get started on a search. And that's probably where I use that kind of stuff the most." |
| Q23 | P10 | "So I would input, I would just say I need an email written to the executives to. A professional email to executives inviting them to this event and yada yada. And so that's how I used it at work most of the time, was to make myself sound more professional." |
| Q24 | P07 | "For Valentine's Day, I was having [my partner's] car detailed inside and out, and I wanted to make him a card with, like, a riddle. And so it kind of like, gave me a riddle for him to start to guess. And I said, can you make it a little bit longer? And then it made it a little bit longer. And then I said, do you have an alternate version? I was just curious. And then they gave me an alternate version. And I must have went with that or the one before it. And it was just like, you know, I just wanted something silly because, like I wasn't going to be able to give him the car detail that day." |
| Q25 | P08 | "I was talking to [a female-coded chatbot] about something, and then she referenced some sort of event that I had taken part in. And she asking me about it, but the only reference to it was some random Instagram post or flyer that came from the event. And so it was clear that she was just drawing from information that Instagram knew about me. But it didn't feel like it was a normal interaction. It wasn't like somebody that would just walk up to you randomly, not knowing you and be like, 'So how was it? Did you like this?' It was just weird." |
| Q26 | P11 | "It could be very stupid, it could like just repeat what you told it. Like you said, 'Oh, for this specific type of experiment LSA-XT can be the optimal choice'. Then the next time you asked 'Hey, do you have like new idea about this assay? Do you think I should use a different type of technology for this assay?' but it will keep mentioning LSA-XT. Because it appears you were the only one who told ChatGPT that LSA-XT is a good technique for this experiment. So it just keeps telling you this is the best." |
| Q27 | P02 | "We were looking for information on doctors who had been involved in nuclear medicine in Canada by decade. The results that kept coming up when we were using generative AI included a lot of people who were physicists, not [medical] doctors, but because doctors ahead of their name. The generative AI just doesn't do that. I am always looking for something that's going to speed up my work or enable me to do things faster, but it has to be accurate. And a lot of this comes in to thinking through what generative, like what the generative AI tools are particularly good at." |
| Q28 | P05 | "And then if it made sense that the patient had improving urine output toward the end of their hospitalization, because we gave diuretics, and I said that we gave diuretics that should have improved urine output. All that the EMR [electronic medical record] ever mentioned, let's say, was that overall fluid status decreased, the in and out balance showed a decrease in fluid status over the last three days of hospitalization. Urine output was never specifically mentioned. It would say, 'during the last three days, urine output increased'. Right. But it didn't. And so while that seems really small, that's actually not cool, because you could have insensate losses that could have been lost through emesys, could have been lost through procedures like a paracentesis, that kind of thing. So it would hallucinate those little tiny details that made the narrative more attractive and made it medically line up more nicely. But if that isn't what actually happened, it should not be in there." |

| Q29 | P13 | "Sometimes [Gemini is] way off and I ... try redirecting it. Then I'll have to make a decision. Do I want to redirect it from what it gave me? OK. It's a middle-aged male. You're giving me pediatric answers and then it will reassess and I'll kind of give it a chance that way. I'll try to just give it more whatever information I have to try to guide into something else, and often it will like click into a more reasonable differential diagnosis that I can go research on real tools. That doesn't happen too often." |
|-----|-----|---|
| Q30 | P14 | "You know for example there's always something about a logo that isn't just quite right done by AI. So my job is to make it human. It can't read and see all those little tricky things like balance and motion and all these intangible ways artists describe things. It does a great job out-of-the-box, I just find the final bringing it home is something that I've needed to do. I sort of coalesce all the ideas that are generated and then redo it myself. I could be cruising and making fifty different alterations with the AI and then I'll grab what I have and then just start picking them apart and assembling and putting them together and seeing what I'm trying to gain in my mind's eye, then do it on Photoshop after." |
| Q31 | P15 | "It's decent enough because it gives you like a really quick and dirty output. But none of these assets actually end up getting used as a final asset, because the output from them is not in a usable stage. There are multiple iterations which happen in our pipeline where it goes from a generated asset to then getting re-touched, re-modeled, re-meshed and then passed down towards other [work] streams. Definitely for IP reasons, I mean it could end up being a huge copyright infringement thing, which makes big problems for the studios. " |

**Table 11: Selected quotes from participants.**